\documentclass[preprint,1p]{elsarticle}
\usepackage{a4wide}
\usepackage{amsmath}
\usepackage{amsfonts}
\usepackage{amssymb}
\usepackage{natbib}
\usepackage{ifpdf}
\usepackage{float}

\newcommand{\be}{\begin{equation}}
\newcommand{\ee}{\end{equation}}
\newcommand{\ba}{\begin{eqnarray}}
\newcommand{\ea}{\end{eqnarray}}
\newcommand{\bc}{}

\newcommand{\bea}{\begin{eqnarray}}
\newcommand{\eea}{\end{eqnarray}}
\newcommand{\nn}{\nonumber}

\newcommand{\m}{\mathcal}
\newcommand{\mbb}{\mathbb}
\newcommand{\mf}{\mathfrak}
\newcommand{\der}{\mathrm{d}}

\biboptions{sort&compress}

\begin{document}

\begin{frontmatter}
\title{Local quanta, unitary inequivalence, and vacuum entanglement}
\author{Mat\'ias Rodr\'iguez-V\'azquez}\ead{mrodvaz@gmail.com}
\author{Marco del Rey}\ead{marco.del.rey@iff.csic.es}
\author{Hans Westman} \ead{hwestman74@gmail.com}
\author{Juan Le\'on\corref{cor1}}\ead{juan.leon@csic.es}
\cortext[cor1]{Corresponding author}
\address{Instituto de F\'isica Fundamental, CSIC, Serrano 113-B, 28006 Madrid, Spain}
\begin{abstract}
In this work we develop a formalism for describing localised quanta for a real-valued Klein-Gordon field in a one-dimensional box $[0, R]$. We quantise the field using {\em non-stationary local modes} which, at some arbitrarily chosen initial time, are completely localised  within the left or the right side of the box. In this concrete set-up we directly face the problems inherent to a notion of local field excitations, usually thought of as elementary particles. Specifically, by computing the Bogoliubov coefficients relating local and standard (global) quantizations, we show that the local quantisation yields a Fock space $\mf F^L$ which is {\em unitarily inequivalent} to the standard one $\mf F^G$. In spite of this, we find that the local creators and annihilators remain well defined in the global Fock space $\mf F^G$, and so do the local number operators associated to the left and right partitions of the box. We end up with a useful mathematical toolbox to analyse and characterise local features of quantum states in $\mf F^G$. Specifically, an analysis of the global vacuum state $|0_G\rangle\in\mf F^G$ in terms of local number operators shows, as expected, the existence of entanglement between the left and right regions of the box. The local vacuum $|0_L\rangle\in\mf F^L$, on the contrary, has a very different character. It is neither cyclic nor separating and displays no entanglement. Further analysis shows that the global vacuum also exhibits a distribution of local excitations reminiscent, in some respects, of a thermal bath. We discuss how the mathematical tools developed herein may open new ways for the analysis of fundamental problems in local quantum field theory.
\end{abstract}
\begin{keyword}
Local Quantum Theory, Vacuum Entanglement, Localizability, Reeh-Schlieder Theorem, Quantum Steering, Local States, Unitary Inequivalence
\end{keyword}
\end{frontmatter}
\newpage
\tableofcontents
\newpage
%
\section{Introduction}
Quantum Field Theory (QFT in short) has proven to be one of the most successful theories in Physics. Its potential to describe the properties of elementary particles has been richly demonstrated within the framework of the Standard Model of Particle Physics. The extraordinary agreement between theoretical and experimental values of the muon $g-2$ anomaly \cite{Blum:2013}, or the recent experimental success vindicating the Higgs mechanism after decades of search \cite{MoriondATLAS,MoriondCMS}, are just two examples among many.

Elementary particles in modern physics are commonly thought of as small localised entities moving around in space. A careful examination, however, reveals such an interpretation to be problematic: in QFT a free particle is represented by a superposition of positive-frequency complex-valued modes which satisfy some field equation (e.g. the Klein-Gordon equation). Yet, no superposition of positive-frequency modes can be localised within a region of space, even for an arbitrarily small period of time \cite{Hegerfeldt:1998}. 

This confusing issue is sometimes mistaken as superluminality, see \cite{Prigogine} for a clarification. In fact, it can be shown that the time derivative $\dot\psi$, for any wave-packet $\psi$ composed exclusively out of positive frequency modes, is non-zero almost everywhere in space.\footnote{One way of seeing this is by noting that positive frequency solutions also satisfy the square root of the Klein-Gordon equation, i.e. the  Schr\"odinger equation $i\dot\phi(\vec{x},t)=\sqrt{-\nabla^2 +m^2} \phi(\vec{x},t)$. From there,  using the {\em antilocality} property of the operator $\sqrt{-\nabla^2 +m^2}$, it follows that the time derivative $\dot\phi$ is necessarily non-zero almost everywhere in space \cite{Masuda}.} For that reason, even if $\psi$  propagates in a perfectly causal manner according to the Klein-Gordon equation, it can hardly represent a localised entity. It is problematic to think of the fundamental field excitations of QFT as `particles' in any common sense of the word.

The problem of localisation can be analysed from other angles, for example in terms of {\em localisation systems}. These are defined in terms of a set of projectors $E_\Delta$ on bounded spatial regions $\Delta$ whose expectation values yield the probability of a position measurement to find the particle within $\Delta$. A theorem by Malament \cite{Malament} shows that in a Minkowski spacetime, under reasonable assumptions for the projector algebra, no such non-trivial set of projectors exists. There is also a general result (valid for both, relativistic or non-relativistic cases) due to Hegerfeldt \cite{Hegerfeldt:1998} proving that, assuming a Hamiltonian with spectrum bounded from below, the expectation value of those projectors is non-zero for almost all times. In particular this applies also to states naively thought to be localised. Also along this line, but in order to describe unsharp localisation systems, Busch \cite{Busch:1999} replaced the use of projectors by more general operators, "effects" (or Positive-Operator Value Measures -- POVM), showing that it is impossible to localise with certainty a particle in any bounded region of space. Furthermore, completing the collection of no-go theorems, Clifton and Halvorson \cite{HalvorsonThesis} have shown, under a set of natural requirements, that it is not possible to define local number operators associated to any finite region of space. At this point it is also worthwhile mention the well-known problems of other efforts, based on the use of putative observables such as the Newton-Wigner position operator \cite{NewtonWigner,Fleming:2000,Halvorson}.

In addition, there is also a different notion of localisation called {\em strict localisability} \cite{Knight:1961,Licht:1963}. The basic idea is that a state, localised within a region of space at some specific moment in time, should be such that the expectation value of any operator associated to a spacelike separated region should be the same as in the vacuum. In other words, average values of local operators will depend on the state only if the observation is made in the region where the state is localised. However, as shown by Knight, no finite superposition of $N$-particle states can be strictly localised. Some researchers have adopted the view that the notion of strict localizability is therefore too strong, and suggested that it should be relaxed by allowing for asymptotic  localization, implemented by exponential fall-offs out of the localisation region. This was called {\em essential localization} and proposed as a criterion for deciding whether a QFT could describe particles \cite{Haag:1965}.

Although the results and theorems discussed above are well-understood mathematically, they nevertheless remain puzzling from a physical point of view, as they indicate that the quanta of QFT are not, at the fundamental level, particles in any common sense of the word. The situation is further complicated when we consider quantum fields in curved spacetimes, or in the presence of an external field, where  there is, in general, no well-defined notion of a particle. This is the well-known {\em particle number ambiguity}, which have led some researchers to claim  that the notion of particle is ultimately not a useful concept. For example, in his book \cite{WaldCurvedBook}, Wald writes: 
\begin{quote}
{\em ``Indeed, I view the lack of an algorithm for defining a preferred notion of `particles' in QFT in curved spacetime to be closely analogous to the lack of an algorithm for defining a preferred system of coordinates in classical general relativity. (Readers familiar only with presentations of special relativity based on the use of global coordinates might well find this fact to be alarming.) In both cases, the lack of an algorithm does not, by itself, pose any difficulty for the formulation of the theory.''} R. Wald
\end{quote}
We shall not be concerned in  this paper with the usefulness of the particle concept in QFT. We will rather make practical use of this ambiguity to provide a non-standard quantisation procedure yielding a QFT which, by construction, contains strictly localised one-particle states.

Our approach can be viewed as a modification and further elaboration on a previous work by Colosi and Rovelli \cite{Rovelli:2009}. Instead of quantizing the field using the standard stationary modes, we employ {\em non-stationary} modes which are, together with their time-derivatives, completely localised within a region of space at some arbitrary chosen time. These modes then evolve freely and spread out to become completely de-localised. The associated creation and annihilation operators can then be used to construct a local Fock space $\mf F^L$ which is distinct from the Fock space $\mf F^G$ associated with the standard quantisation based on the global (i.e. non-localisable) stationary modes. 

The local quantisation brings along a notion of {\em strictly localised particle states} which means that one or more assumptions of the theorems and results discussed above do not hold in our construction. Intriguingly, the local Fock space $\mf F^L$ can be shown to be {\em unitarily inequivalent} to the global Fock space $\mf F^G$. This could be taken as an indication that the local quantization, and the associated localised particle states, are problematic. However, they yield a self-consistent QFT with well-defined state evolution and quanta having a well-defined energy expectation value after the relevant local vacuum energy has been subtracted.

This paper is organised as follows. Section \ref{sec:Background} serves to fix notation and conventions as well as to provide the basic background material. In particular, we make explicit the arbitrariness of the choice of a complete set of orthonormal modes for the quantisation procedure. In Section \ref{sec:NonUniqueness} we briefly discuss the standard quantisation based on stationary modes yielding the standard Fock space $\mf F^G$. We then discuss the relationship between quantum theories obtained by different choices of modes and provide a sufficient condition for unitary inequivalence. In Section \ref{sec:LocalFormulation}  a new set of local modes is introduced in order to construct the local Fock space $\mf F^L$. Later, in Section \ref{dealunineq}, we prove that the local and the global representations, are unitarily inequivalent. In Section \ref{locstates} we show the local one-particle states are {\em strictly} localised and evolve causally. We also prove that the Hamiltonian can be regularised by subtraction of the local vacuum energy. By showing in Section \ref{locanal} that the local creators and annihilators are well-defined operators in the global Fock space $\mf F^G$, we end up with a mathematical toolbox enabling us to analyse and characterise states in $\mf F^G$. We later check the properties of the vacuum in terms of local number operators. We exhibit the expectation values of the local number operators and quantify their correlations between the two regions. We also introduce a set of quasi-local states on $\mf F^G$. In the section \ref{sec:quasilocal} we study the properties of these quasi-local states, including the positivity of energy and their failure to be strictly localised, while comparing them to local and global states. Then we discuss the possibility of quantum steering using the vacuum and how it relates to the Reeh-Schlieder theorem. We end up with an outline of future extensions of this work and a summary of the conclusions. 

\section{Background material, notation, and conventions}
\label{sec:Background}
In this section we shall review some background material while fixing notations and conventions used throughout this paper. 
\subsection{Classical scalar field}
\label{sec:ClassicalScalarField}
Consider a free real scalar field $\phi(x,t)$ in a one dimensional cavity of size $R$. Varying the Klein-Gordon action
\begin{align}
S=\frac{1}{2}\int \der x \left(\eta^{\mu\nu}\partial_\mu \phi\partial_\nu \phi-\mu^2\phi^2\right),
\end{align}
and imposing Dirichlet boundary conditions $\phi(0,t)=\phi(R,t)=0$, we obtain the Klein-Gordon equation
\begin{align}
\label{eqn:KG3D}
\partial_\mu\partial^\mu \phi+\mu^2\phi=(\Box + \mu^2)\phi(x,t)=0,
\end{align}
where we have put $\hbar=c=1$ and  $\eta_{\mu\nu}=diag(+1,-1)$. The linearity of the equation implies that the space of solutions forms a vector space $\mathfrak S$. 
\subsection{Klein-Gordon inner product}
The classical field is throughout this paper taken to be real valued $\phi(x,t) : [0, R] \times \mbb R\rightarrow \mbb R$. Nevertheless, at the QFT level, complex valued solutions $\phi: [0, R] \times \mbb R\rightarrow \mbb C$ occur naturally and describe one-particle states. The vector space $\mathfrak S^{\mathbb C}$ of complex valued solutions of \eqref{eqn:KG3D} is equipped with a sesqui-linear (pseudo) inner product called the Klein-Gordon inner product:
\begin{align}\label{eqn:KGinnerprod}
(\phi_1|\phi_2)=i\int_0^R dx\phi_1^*(x,t)\overset{\leftrightarrow}{\partial}_t\phi_2(x,t)=i\int_0^R dx(\phi_1^*(x,t)\dot \phi_2(x,t)-\dot\phi_1^*(x,t)\phi_2(x,t)),
\end{align}
with $\dot{}\equiv\partial_t$. The quantity $(\phi_1|\phi_2)$ is conserved in time only if $\phi_1$ and $\phi_2$ are both solutions and subject to the {\em same} boundary conditions, i.e. $\phi_1,\phi_2 \in \mathfrak{S}^{\mathbb C}$. We note that the Klein-Gordon inner product is not positively definite on $\mf S^{\mbb C}$. Thus, although $\mf S^{\mbb C}$ is a vector space, it is not a Hilbert space. In fact, the Klein-Gordon product partitions the solutions space $\mathfrak S^{\mathbb C}$ into three subsets of solutions:
\begin{align}
\phi\in\mathfrak S_+^{\mathbb C}\quad\Rightarrow\quad (\phi|\phi)>0,\nn\\
\phi\in\mathfrak S_-^{\mathbb C}\quad\Rightarrow\quad (\phi|\phi)<0,\nn\\
\phi\in\mathfrak S_0^{\mathbb C}\quad\Rightarrow\quad (\phi|\phi)=0,\nn\\
\end{align}
corresponding to solutions with positive, negative, and zero Klein-Gordon norm. Real-valued solutions are members of $\mathfrak S_0^{\mathbb C}$. Moreover, neither of the three subsets $\mathfrak S_+^{\mathbb C}$, $\mathfrak S_-^{\mathbb C}$, and $\mathfrak S_0^{\mathbb C}$ form vector spaces, let alone Hilbert spaces.
\subsection{Mode bases and the one-particle Hilbert space}
We can isolate a one-particle Hilbert space by introducing a complete and orthonormal basis, $\{f_m(x,t),f_m^*(x,t)\}$ with $m\in \mathbb N^+$, of the vector space $\mathfrak S^{\mathbb C}$.\footnote{The structure we need in order to isolate a one-particle Hilbert space $\mf H$ in the solution space $\mf S^{\mbb C}$ is a {\em complex structure} \cite{WaldCurvedBook}. In our notation it takes the form $\mf J=i\left( \sum_N|f_m)(f_m|+|f_m^*)(f_m^*|\right)$.} We will require all $f_m(x,t)$ to have positive norm, which implies that the complex conjugate ones $f^*_m(x,t)$ have negative norm. The orthonormality conditions read  
\begin{align}\label{eqn:orthorel}
(f_m|f_n)=\delta_{mn},\quad (f_m^*|f_n^*)=-\delta_{mn},\quad (f_m^*|f_n)=0.
\end{align}
A set of modes form a complete set if for any solution $\phi(x,t)\in\mf S^{\mbb C}$ we have the following identity
\begin{align}
\phi(x,t)=\sum_m(f_m|\phi)f_m(x,t)-(f_m^*|\phi)f_m^*(x,t),
\end{align}
up to a zero measure set of points $x\in[0, R]$. Writing out this identity  using the definition of the Klein-Gordon inner product \eqref{eqn:KGinnerprod} yields
\begin{align}
\phi(x,t)&=i\int dx'\sum_m \left(f_m^*(x',t)f_m(x,t)-f_m(x',t)f_m^*(x,t)\right)\dot\phi(x',t)\nn\\
&\qquad -\left(\dot f_m^*(x',t)f_m(x,t)-\dot f_m(x',t)f_m^*(x,t)\right)\phi(x',t).
\end{align}
Since the Klein-Gordon equation is a second-order partial differential equation, $\phi(x',t)$ and $\dot\phi(x',t)$ are independently specifiable. Thus, for the identity to hold for any solution $\phi$, and at any time $t$, we deduce the following completeness relations
\begin{align}\label{eqn:comprel}
0&=\sum_mf_m^*(x',t)f_m(x,t)-f_m(x',t)f_m^*(x,t),\nn\\
\delta(x-x')&=i\sum_m\dot f_m(x',t)f_m^*(x,t)-\dot f_m^*(x',t)f_m(x,t).
\end{align}
If we restrict ourselves to real fields, any such a field  $\phi(x,t)$ can be expanded as
\begin{align}
\phi(x,t)&=\sum f_m(x,t)a_m+f_m^*(x,t)a_m^*,
\end{align}
where $a_m=(f_m|\phi)$ are complex numbers and $a_m^*=-(f_m^*|\phi)$, the complex conjugates of $a_m$.

The Hilbert space of one-particle states $\mf H$ is then defined to be the vector space spanned by the positive norm modes $f_m$, i.e. 
\begin{align}
\mf H=span(f_m).
\label{eqn:oneparticle}
\end{align}
The Klein-Gordon product, when restricted to the subspace $\mf H\subset\mf S^{\mathbb C}$, is  by construction a positive definite sesqui-linear product. Therefore, $\mf H$ is a Hilbert space. In general $\mf H$ will depend on the choice of basis $\{f_m,f_m^*\}$ as defined in \eqref{eqn:oneparticle},  leading to the well-known {\em particle number ambiguity} in QFT \cite{BirrellDaviesBook}.
\subsection{Dirac notation}
To keep notation tidy and transparent it will be useful to introduce a Dirac notation to denote the vectors of $\mf S^{\mbb C}$. To that end we make the identification $\phi(x,t)\sim |\phi)\in\mf S^{\mbb C}$. We will also consider the dual space $\mf S^{\mbb C*}$; the vector space of linear maps $\mf m:\mf S^{\mathbb C}\rightarrow \mathbb C$. The Klein-Gordon product $(\cdot|\cdot)\rightarrow \mbb C$ associates any vector $|\phi)\in \mf S^{\mbb C}$ to a member of the dual vector space through $(\phi|\cdot)\in\mf S^{\mbb C*}$ and so we will  write $(\phi|\in \mf S^{\mbb C*}$. In this  notation the completeness relations \eqref{eqn:comprel} take the succinct form
\begin{align}
&\sum_m |f_m)(f_m|-|f_m^*)(f_m^*|=1,
\end{align}
where $1$ denotes the identity operator on the vector space of solutions $\mf S^{\mbb C}$. Note that we use `round' brackets $|\phi)$ for vectors in $\mf S^{\mbb C}$. In contrast we will use the standard brackets $|\psi\rangle$ to denote states of the corresponding QFT to which we now turn.
\subsection{Quantization}
\label{sec:GlobalQuantization}
In order to quantise the real-valued classical field $\phi(x,t)$ we first perform the Legendre transformation, which yields the Hamiltonian and canonical momenta
\begin{align} \label{eqn:GlobalHamiltonian}
H=\int dx \frac{1}{2}\left[\pi^2 + (\nabla \phi)^2 +\mu^2\phi^2\right],\qquad \pi=\dot\phi.
\end{align}
Standard Dirac quantisation now requires us to turn $\phi$ and $\pi$ into operators $\hat \phi$ and $\hat\pi$, satisfying equal-time canonical commutation relations
\begin{align}
\label{eqn:CanonicalCommutationRelations}
[\hat \phi(x,t), \hat \pi(y,t)]=i\delta(x-y),\quad [\hat \phi(x,t),\hat \phi(y,t)]=0, \quad [\hat \pi(x,t),\hat \pi(y,t)]=0.
\end{align}
For notational convenience and since no confusion arises, we will refer to these operators from now on as $\phi$ and $\pi$, with the hats `\ $\hat{}$\ '  omitted.

In order to provide a Fock space representation of the commutator algebra \eqref{eqn:CanonicalCommutationRelations} we expand the field in some complete and orthonormal basis $\{f_m,f_m^*\}$ and write
%
\begin{align}
\phi(x,t)&=\sum_m f_m(x,t)a_m+f_m^*(x,t)a_m^{\dagger},\nn\\
\pi(x,t)&=\dot\phi(x,t)=\sum_m \dot f_m(x,t)a_m+\dot f_m^*(x,t)a_m^{\dagger},
\end{align}%
where $\dot f_m\equiv \partial_t f_m$, and $a_m$ and $a_m^\dagger$ have been promoted into operators. If the modes $\{f_m,f_m^*\}$ satisfy the (second of the) completeness relations \eqref{eqn:comprel} then the following standard commutator algebra of creation and annihilation operators
\begin{align}\label{eqn:commrel}
[a_m,a_n^\dagger]=\delta_{mn},\qquad [a_m^{\dagger},a_n^{\dagger}]=0, \quad [a_m,a_n]=0,
\end{align}
ensures that we satisfy  \eqref{eqn:CanonicalCommutationRelations}. As usual, we will define the vacuum state $|0\rangle$ to be the state annihilated by all operators $a_m$, i.e. 
\begin{align}
a_m|0\rangle=0\ \forall m\in \mathbb N^+.
\end{align}
A complete and orthonormal set of basis vectors $|n_1,n_2,\dots\rangle$ of the corresponding Fock space $\mf F$ is obtained by repeated application of the creation operators on the vacuum state:
\begin{align}
\label{eqn:fockbasis}
|n_1,n_2,\dots\rangle=\prod_m\frac{(a_m^\dagger)^{n_m}}{\sqrt{n_m!}}|0\rangle,
\end{align}
where the total number of particles in each basis state is required to be finite, $\sum_{k}  n_k < \infty$, ensuring that $\mf F$ is a separable Hilbert space \cite{HaagBook}. 

$\mf F$ is, as spanned by this basis, nothing but the symmetrised Fock space associated with the bosonic  one-particle Hilbert space $\mf H$, i.e.
\begin{align}
\label{eqn:GlobalFockSpace}
\mf F(\mf H)=\bigoplus_{n=0}^{\infty}\sideset{}{_{S}}\bigotimes^n  \mathfrak{H} =\mathbb{C}\oplus \mathfrak{H} \oplus (\mathfrak{H}\otimes_S\mathfrak{H})\oplus \ldots.
\end{align}
Here we note that the one-particle subspace spanned by the states $|1_m\rangle\equiv a_m^\dagger|0\rangle$ is indeed the same as $\mf H$, or explicitly $f_m(x,t)=\langle0|\phi(x,t)|1_m\rangle$ \cite{Schweber:2005}.
\section{Non-uniqueness of the quantisation procedure}
\label{sec:NonUniqueness}
In the previous section we described how to quantise a classical real-valued Klein-Gordon field, and deliberately kept the choice of orthonormal modes $\{f_k,f_k^*\}$ unspecified. Although this choice does not affect the classical field theory the situation is different at the QFT level. In fact, different choices of modes may lead to {\em unitarily inequivalent} Fock space representations. A well-known example in this regard is of course the Fulling-Rindler quantisation \cite{Fulling:1973}. Examples of a different kind are given in \cite{Lupher}. By the Stone-von-Neumann theorem \cite{VonNeumann:1931} this is something that can happen only for  systems with infinitely many degrees of freedom, which is precisely the case of QFT \cite{Ruetsche}.
\subsection{Standard (global) quantization}\label{globalquantization}
The standard set of complete and orthonormal modes for a quantum field in a cavity is given by the normal modes
\begin{align}
\label{eqn:GlobalMode}
U_N(x,t)=\m U_N(x)e^{-i\Omega_Nt}=\frac{1}{\sqrt{R\Omega_N}}\sin \frac{\pi N x}{R}e^{-i\Omega_Nt},\quad U_N^*(x,t)=\m U_N(x)e^{+i\Omega_Nt},
\end{align}
with $\Omega_N^2=\frac{\pi^2 N^2}{R^2}+\mu^2$. We note that  $\{U_N,U_N^*\}$ are all stationary solutions with the time dependence confined to a complex phase $e^{\pm i\Omega_Nt}$. By computing the Klein-Gordon inner products, e.g. $(U_N|U_M)$, it is easily checked that these modes satisfy the orthogonality conditions \eqref{eqn:orthorel}. That they form a complete set of modes, and so satisfy the completeness relations \eqref{eqn:comprel}, follows from the fact that they are stationary: the first of the completeness relations is identically satisfied, while the second one is satisfied because of the identity of Fourier analysis 
\begin{align}
\sum_N\frac{2}{R}\sin \frac{N\pi x}{R}\sin \frac{N\pi x'}{R}=\delta(x-x').
\label{eqn:compglob}
\end{align}
Thus we have: 

\begin{equation}
\sum_N |U_N)(U_N|-|U_N^*)(U_N^*| = 1.
\end{equation}

With this choice of modes, the field operator $\phi$ and its conjugate momentum $\pi$ take the form
\begin{align}
\phi(x,t)&=\sum_N U_N(x,t)A_N+U^*_N(x,t)A_N^{\dagger},\nn\\
\pi(x,t)&=\sum_N-i\Omega_N\left(U_N(x,t)A_N-U^*_N(x,t)A_N^{\dagger}\right).
\end{align}%
Now, by making use of the commutation relations \eqref{eqn:commrel}, a very simple expression of the (regularised) Hamiltonian operator can be obtained
\begin{align}\label{eqn:globalhamiltonian}
H^G=H-\langle0_G|H|0_G\rangle=\sum_N \Omega_NA_N^\dagger A_N,
\end{align}
where the infinite vacuum energy $\langle0_G|H|0_G\rangle=\sum_N\frac{1}{2}\Omega_N$ has been removed. The state $|0_G\rangle$ annihilated by all $A_N$, will be referred to hereafter as the {\em global vacuum}. The basis vectors of the corresponding \emph{global} Fock space, denoted by $\mf F^G$, are then
\begin{align}
\label{eqn:globfockbasis}
|n_1,n_2,\dots\rangle=\prod_N\frac{(A_N^\dagger)^{n_N}}{\sqrt{n_N!}}|0\rangle,
\end{align}
and correspond to energy eigenstates of the Hamiltonian $H^G$. Needless to say, the usefulness of the global modes \eqref{eqn:GlobalMode} stems from the fact that they diagonalise the Hamiltonian operator.

We call the basis \eqref{eqn:GlobalMode} a {\em global} basis, since no state in the corresponding one-particle Hilbert space $\mf H^G=span(U_N)$ can be fully contained within a subregion of $[0, R]$ for any arbitrarily small time interval $\Delta t$. As follows from a theorem by Hegerfeldt \cite{Hegerfeldt:1998}, there is no state such that $\phi(x,\tau)=\dot{\phi}(x,\tau)=0$ for all $r<x<R$ at any  time instant $t=\tau$. Instead, all states in $\mf H^G$  have, at almost all time, support in the entire cavity, i.e. they are \textit{global}. As a matter of fact, the non-localizability of one-particle states in Minkowski spacetime is well-known and it has been noted and widely studied in several works, e.g. \cite{Knight:1961, Licht:1963,Prigogine}.
\subsection{Positive norm vs positive frequency}
It is important to stress that in the standard global quantisation the {\em positive (negative) norm modes} coincide with {\em positive (negative) frequency modes}; two conceptually distinct notions, which should not be confused. In fact, what is important for the quantisation procedure and the construction of a Fock space is not the partitioning of modes into positive and negative frequencies, but rather the partitioning into positive and negative norm modes. The latter notion does not require the basic field equations to admit symmetry under time translations but generalizes straightforwardly to non-stationary equations such as a quantum field in a time-dependent spacetime, or in the presence of a varying external field. This is so since the Klein-Gordon inner product, which defines the partitioning into positive and negative norm solutions, remains well defined also in these situations.

We shall exploit this fact in the next section. 
\subsection{Bogoliubov transformations}
\label{sec:BogoliubovTransformations}
Let us  explore here the relationship between quantizations based on different choices of modes. To this end, let $\{f_m,f^*_m\}$ and $\{\tilde f_m,\tilde f^*_m\}$ be two complete sets of orthonormal modes. Then we can expand the quantum field in two distinct ways:

\begin{align}
\phi(x,t)=&\sum_m f_m(x,t)a_m+f_m^*(x,t)a_m^\dagger=\sum_m \tilde f_m(x,t)\tilde a_m+\tilde f_m^*(x,t)\tilde a_m^\dagger.
\end{align}
Using the orthogonality relations \eqref{eqn:orthorel} we can immediately read off the relations 
\begin{align}
\label{eqn:modes}
\tilde a_m&=\sum_n(\tilde f_m|f_n)a_n+(\tilde f_m|f_n^*)a_n^\dagger,\\
\tilde a_m^\dagger&=\sum_n(f_n|\tilde f_m)a_n^\dagger+(f_n^*
|\tilde f_m)a_n.
\end{align}
The complex coefficients $(\tilde f_m|f_n)$, $(\tilde f_m|f_n^*)$, $(f_n|\tilde f_m)$, and $(f_n^*|\tilde f_m)$ are the \emph{Bogoliubov coefficients}\footnote{More formally speaking, a Bogoliubov transformation is a transformation that preserves the symplectic structure in the case of classical fields, or the canonical commutation relations in a QFT. }. In the literature they   are commonly denoted by $\alpha$ and $\beta$ (and its complex conjugate), defined by $\tilde f_m=\sum_n \alpha_{mn}f_n+\beta_{mn}f_n^*$ so that
\begin{align}
\label{eqn:GenericBogosAlphaBeta}
\alpha_{mn}\equiv(f_n|\tilde f_m),\qquad \beta_{mn}\equiv-(f_n^*|\tilde f_m).
\end{align}
\subsection{A sufficient condition for unitary inequivalence}
We say that two Fock space representations are unitarily equivalent if there exists a unitary map $\mf B:\mf F\rightarrow \tilde{\mf F}$ that relates the  Fock spaces associated with the representations, $\mf F$ and $\tilde{\mf F}$. Necessary and sufficient conditions for two Fock space representations to be unitarily equivalent  are given in \cite{WaldCurvedBook}.

In this paper we shall demonstrate unitary {\em inequivalence} of two Fock space representations and will therefore only need the following  condition.

\begin{quote}
{\bf Sufficient condition for unitary inequivalence:} Two Fock-space representations are unitarily inequivalent if the vacuum state of one representation has infinitely many particles in terms of the number operator of the other representation, i.e. 
\begin{align}\label{eqn:unineqcond}
\sum_m\langle\tilde 0|N_m|\tilde 0\rangle=\sum_m\langle0|\tilde N_m|0\rangle=\sum_{mn}|(\tilde f_m|f^*_n)|^2=\infty,
\end{align}
where $a_m|0\rangle=\tilde a_m|\tilde 0\rangle=0\ \forall m\in\mbb N^+$, $N_m\equiv a_m^\dagger a_m$, and $\tilde N_m\equiv \tilde a_m^\dagger\tilde a_m$.
\end{quote}
Well-known cases of unitarily inequivalent representations can be found in \cite{Takagi:1986, Fulling:1973, Lupher, Haag:1965}. In this paper we shall provide a new example.
\section{Quantisation based on local non-stationary modes}
\label{sec:LocalFormulation}
The elementary excitations of the field, defined by the Fock quantisation described in Section \ref{globalquantization}, consist of global modes which are also stationary. As already mentioned before, using only positive frequencies it is not possible to construct wave packets that are completely localised within a subregion $\mf R\subset[0, R]$ of the cavity in the sense that $\phi(x,t)=\dot\phi(x,t)=0$ if $x\notin\mf R$. This feature is a consequence of Hegerfeldt's theorem \cite{Hegerfeldt:1998}. Forcing $\phi=0$ outside the region $\mf R$ implies a non-zero $\dot\phi$ outside the subregion resulting in a wave-packet that at an infinitesimal time later would become non-zero almost everywhere outside the subregion of localisation. For such a case, the Hamiltonian density would be non-zero outside the subregion, and in this sense, states in $\mf H^G$ cannot be localised.

The standard quantisation of a free field relies on global non-localised excitations. Given the freedom in the choice of modes when quantizing a field (see Section \ref{sec:GlobalQuantization}) it is suggestive to try, alternatively, to quantise the scalar field using modes representing local excitations. Such an excitation would be, at some instant $t=\tau$, localised and hereafter free to evolve and causally spread out. 

These local modes can then be used to find a Fock space representation of the canonical commutation relations as outlined previously, and a `local' Fock space $\mf F^L$ which hopefully admits strictly localised one-particle states. Nevertheless, as will be demonstrated in Section \ref{dealunineq}, the local Fock space $\mf F^L$ will turn out to be unitarily inequivalent to $\mf F^G$.
\begin{figure}[H]
  \centering
  \includegraphics[width=\textwidth]{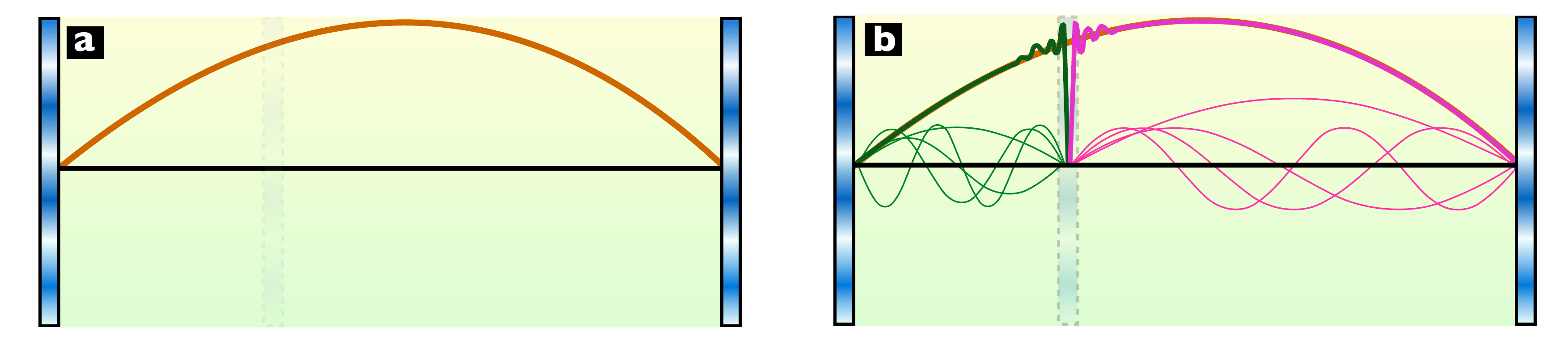}      \caption{a) Simple scheme for quantisation in a cavity. Global modes are used to define the one-particle Hilbert space. b) The local modes (defined by imagining an instant partitioning of the cavity), can be used to define a local one-excitation space. They form a complete set of modes, which can expand the global modes almost everywhere. In particular we see in the figure how a decomposition in local modes (up to a cutoff) would look for a global mode $N=1$ at $t=0$. }
\label{fig:cavity}
\end{figure}
\subsection{Defining a new set of local modes}
In order to motivate the form of the local modes we consider what happens if we place a perfect mirror at $x=r$,  imposing a Dirichlet boundary condition at that point, $\phi(x=r,t)=0,\ \forall t\in\mbb R$. Mathematically speaking we now have two distinct cavities, each with a quantum field. The complete set of orthonormal modes, $\{v_l(x,t),v_l^*(x,t)\}$ and $\{\bar v_l(x,t),\bar v_l^*(x,t)\}$ for the left and right cavities respectively, are taken to be the usual stationary modes
\begin{align}\label{eqn:Dlocalmodes}
v_l(x,t)&=\frac{1}{\sqrt{r\omega_l}}\sin\frac{l\pi x}{r}e^{-i\omega_l t},\qquad 
v_l^*(x,t)=\frac{1}{\sqrt{r\omega_l}}\sin\frac{l\pi x}{r}e^{+i\omega_l t},\nn\\
\bar v_l(x,t)&=\frac{1}{\sqrt{\bar r\bar \omega_l}}\sin\frac{l\pi (x-r)}{\bar  r}e^{-i\omega_l t},\qquad 
\bar v_l^*(x,t)=\frac{1}{\sqrt{r\bar \omega_l}}\sin\frac{l\pi (x-r)}{\bar r}e^{+i\bar \omega_l t},
\end{align}
where $\omega_l^2=\frac{\pi^2 l^2}{r^2}+\mu^2$, and $\bar\omega_{l}^2=\frac{\pi^2 l^2}{\bar r^2}+\mu^2$, with  $\bar{r}=R-r$. We now quantise the two systems yielding two quantum fields in two distinct cavities. The Fock spaces of the quantum excitations for each cavity are, by construction, localised  within $[0, r]$ and $[r, R]$, respectively. 

We could now try to analyse the quantum field in the {\em entire} cavity $[0, R]$ using such local excitations. It is clear that the introduction of a mirror at $x=r$ necessarily changes the physical conditions and we therefore are no longer dealing with the same physical system, i.e. the original cavity in $[0, R]$. At the mathematical level, the introduction of the Dirichlet boundary condition  changes the solution space to something different than $\mf S^{\mbb C}$. Specifically, the modes $\{v_l(x,t),v_l^*(x,t)\}$ and $\{\bar v_l(x,t),\bar v_l^*(x,t)\}$ no longer form a basis for $\mf S^{\mbb C}$. For this reason, modes of this type are not appropriate for quantizing the field of the full cavity $[0, R]$.

The remedy, however, is simple: instead we will use the local modes \eqref{eqn:Dlocalmodes} to define the Cauchy {\em initial conditions}. Although we could take modes well localised at different moments in time, we shall only consider here, for simplicity, modes $\{u_l,u_l^ *\}$ and $\{\bar u_l,\bar u_l^{ *}\}$ localised at time $t=0$ . These modes are then free to spread out over the entire box $[0, R]$ with no Dirichlet boundary condition imposed at $x=r$. This guarantees that they are still members of the complex solution space, i.e. $u_l,u_l^{*},\bar u_l,\bar u_l^{ *}\in\mf S^{\mbb C}$. 


In order to mimic the local modes we simply read off the initial conditions from the modes \eqref{eqn:Dlocalmodes} evaluated at $t=0$. This yields,
\begin{align}\label{eqn:locmod}
u_l(x,t=0)&=\frac{\theta(r-x)}{\sqrt{r\omega_l}}\sin\frac{l \pi x}{r}=\chi_l(x),\qquad \dot u_l(x,t=0)=-i\omega_l\chi_l(x),\nn\\
\bar u_l(x,t=0)&=\frac{\theta(x-r)}{\sqrt{\bar r\bar \omega_l}}\sin\frac{l \pi (x-r)}{\bar r}=\bar \chi_l(x),\qquad \dot{\bar u}_l(x,t=0)=-i\bar \omega_l\bar \chi_l(x).
\end{align}
%

Before we determine the form of the local modes $\{u_l(x,t),u_l^*(x,t)\}$ and $\{\bar u_l(x,t),\bar u_l^*(x,t)\}$ for an arbitrary time $t$, i.e. solve the Cauchy problem, we should make sure that they do indeed provide a complete and orthonormal basis for the complex solutions space $\mf S^{\mbb C}$. Indeed, by explicit calculation (conveniently done at the specific time $t=0$)  we can verify that
\begin{align}
(u_m|u_l)=\delta_{ml},\qquad (u_m^*|u_l^*)=-\delta_{ml}, \qquad (\bar{u}_m^*|\bar{u}_l^*)=-\delta_{ml},\qquad ( \bar u_m|\bar u_)=\delta_{ml}.
\end{align}

That the modes form a complete set of solutions for $\mf S^{\mbb C}$ can be seen as follows. First we note that at time $t=0$ the modes coincide with the Fourier basis on $[0,  r]$ and $[r, R]$. By Carleson's theorem of Fourier analysis \cite{Carleson:1966} we  have pointwise convergence for almost all points $x\in[0,  R]$, i.e. we have convergence in $L^2([0, R],\mbb C)$ norm.\footnote{We note that if the field $\phi$ is expanded using the local modes, its value in that mode basis at $x=r$ at time $t=0$ is identically zero. Thus, we cannot expect to have convergence at $x=r$. Nevertheless, for almost all other points in $[0, R]$ we will have pointwise convergence.} This means that we can generate any initial conditions at $t=0$ (up to equivalence in $L^2([0, R],\mbb C)$ norm) and thus any solution of $\mf S^{\mbb C}$ (Check Figure \ref{fig:cavity} for an illustration).  By relating the local modes to the global ones through the Bogoliobov transformations and using the well-known completeness properties for the latter, one can also show that the local modes satisfy \eqref{eqn:comprel} for an arbitrary time $t$. Hence, in Dirac notation we  have
\begin{align}
\sum_l |u_l)(u_l|+|\bar u_l)(\bar u_l|-|u_l^*)( u_l^*|-|\bar u_l^*)( \bar u_l^*|=1.
\end{align}
\subsection{Bogoliubov coefficients and evolution}
\label{subsec:BogoliubovCoef}
\begin{figure}[H]
  \centering
    \includegraphics[width=\textwidth]{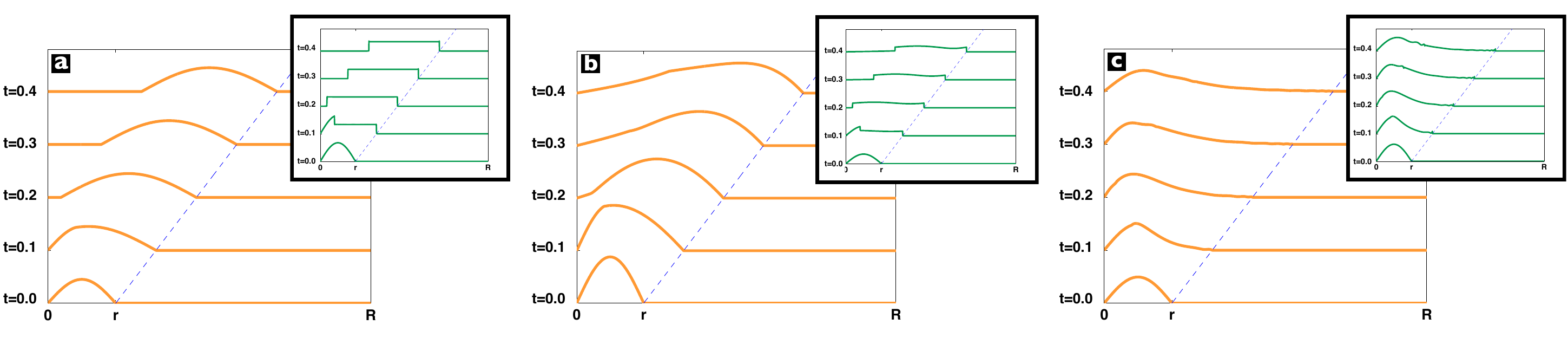}
      \caption{Evolution of the first-excited local mode $u_{m=1}(x,t)$ for different times $t=0,0.1R \ldots 0.5R$. The mode is localised at t=0 in $\mf R =[0, 0.21R] $ within a cavity of size $R$ . The blue dashed line represents the light-cone. a) Massless case. b) Same but with $\mu=1/r=1/(0.21R)$. c) Same but with $\mu=5/r=5/(0.21R)$. 
 We can verify that, after the localisation event in $\mf R$  at $t=0$ the elementary excitation \emph{causally spreads out}, and so does its time derivative $\dot u_{k=1}(x,t)$. The mixing of both positive and negative global frequencies has allowed us to build up a localised mode avoiding  the non-causal infinite tails that Hegerfeldt's theorem would imply.}
\label{fig:UEvolution}
\end{figure}

In order to obtain the modes $u_m(x,t)$ and $\bar u_m(x,t)$ for any time $t$ we simply make use of the completeness property  \eqref{eqn:compglob}:
\begin{align}
|u_m)&=\left(\sum_N |U_N)(U_N|-|U_N^*)(U_N^*|\right)|u_m)=\sum_N (U_N|u_m)|U_N)-(U_N^*|u_m)|U_N^*),\nn\\
|\bar u_m)&=\left(\sum_N |U_N)(U_N|-|U_N^*)(U_N^*|\right)|\bar u_m)=\sum_N (U_N|\bar u_m)|U_N)-(U_N^*|\bar u_m)|U_N^*),
\end{align}
or equivalently 
\begin{align}\label{eqn:localmodfromglobal}
u_m(x,t)&=\sum_N (U_N|u_m)U_N(x,t)-(U_N^*|u_m)U_N^*(x,t),\nn\\
\bar u_m(x,t)&=\sum_N (U_N|\bar u_m)U_N(x,t)-(U_N^*|\bar u_m)U_N^*(x,t).
\end{align}
The Bogoliubov coefficients, $(u_m|U_N)$, $(u_m|U^*_N)$, etc., are independent of which time $t$ we calculate them. Indeed, they can be conveniently calculated by easily taking $t=0$ and using the relations \eqref{eqn:localmodfromglobal}. A straightforward calculation then yields
\begin{subequations}
\label{eqn:BogoliubovsCoefficients}
\begin{align}
(u_m|U_N)&=(\omega_m+\Omega_N)\m V_{mN},\nn\\
(u_m|U_N^*)&=(\omega_m-\Omega_N)\m V_{mN}, \nn\\
(\bar u_m|U_N)&=(\bar \omega_m+\Omega_N)\bar{ \m V}_{mN}, \nn\\
(\bar u_m|U_N^*)&=(\bar \omega_m-\Omega_N)\bar{ \m V}_{mN}, 
\end{align}
\end{subequations}
where
\begin{align}
\m V_{mN}&=\int_0^R dx \m U_N(x)\chi_m(x)=\frac{1}{\sqrt{Rr\Omega_N\omega_m}}\frac{\frac{m\pi}{r}(-1)^m}{\Omega_N^2-\omega_m^2}\sin\frac{N\pi r}{R},\\
\bar{\m V}_{mN}&=\int_0^R dx \m U_N(x)\bar \chi_m(x)=-\frac{1}{\sqrt{R\bar r\Omega_N\bar\omega_m}}\frac{\frac{m\pi}{\bar r}(-1)^{m+N}}{\Omega_N^2-\bar\omega_m^2}\sin\frac{N\pi r}{R}.
\end{align}
Using \eqref{eqn:localmodfromglobal} we can see that the local modes at any time $t$ are given by:
\begin{align}\label{modexp}
u_m(x,t)&=\sum_N\left((\omega_m+\Omega_N)e^{-i\Omega_Nt}-(\omega_m-\Omega_N)e^{i\Omega_Nt}\right) \m V_{mN}  \m U_N(x),\nn\\
\bar u_m(x,t)&=\sum_N\left((\bar \omega_m+\Omega_N)e^{-i\Omega_Nt}-(\bar \omega_m -\Omega_N)e^{i\Omega_Nt}\right)  \bar{\m V}_{mN} \m U_N(x).
\end{align}
Although it is not manifest from the form of the mode expansions \eqref{modexp}, at $t=0$ the local modes $u_m$ and $\bar u_m$ and their time derivatives $\dot u_m$ and $\dot{\bar u}_m$ are zero outside their respective region of localisation. Furthermore, the local modes  $u_k(x,t)$ and $\bar u_k(x,t)$ and their time-derivatives spread out causally from the initial region . This is illustrated for the first-excited mode $u_{m=1}$ in Figure \ref{fig:UEvolution}.
\subsection{Local quantization}
We now turn to the quantisation using these local modes. First we expand the field operator $\phi(x,t)$ using the local modes 
\begin{align}
\phi(x,t)=\sum_m u_m(x,t)a_m+\bar u_m(x,t)\bar a_m+u_m^*(x,t)a_m^\dagger+\bar u^*_m(x,t)\bar a_m^\dagger.
\end{align}
The expressions relating the local and global annihilators are given by
\begin{align}
a_m&=\sum_N (u_m|U_N)A_N+(u_m|U_N^*)A_N^\dagger,\qquad a_m^\dagger=\sum_N (U_N|u_m)A_N^\dagger+(U_N^*|u_m)A_N,\nn\\
\bar a_m&=\sum_N (\bar u_m|U_N)A_N+(\bar u_m|U_N^*)A_N^\dagger,\qquad \bar a_m^\dagger=\sum_N (U_N|\bar u_m)A_N^\dagger+(U_N^*|\bar u_m)A_N.
\label{eqn:afromA}
\end{align}
The commutation relations
\begin{align}
[a_m,a_n]=0,\quad [a_m, a_n^\dagger]=\delta_{mn},\quad [\bar a_m,\bar a_n]=0,\quad [\bar a_m, \bar a_n^\dagger]=\delta_{mn}, \quad [a_m,\bar a_n]=0,\quad [a_m, \bar a_n^\dagger]=0,\nn
\end{align}
and their Hermitian conjugates ensure that the canonical commutation relations \eqref{eqn:CanonicalCommutationRelations} are satisfied. Besides, the local vacuum state $|0_L\rangle$ is defined as the state annihilated by both $a_m$ and $\bar a_m$
\begin{align}
a_m|0_L\rangle=\bar a_m|0_L\rangle=0\ \forall m\in\mbb N^+.
\end{align}
The orthonormal basis vectors are given, as usual, by the repeated application of the creation operators
\begin{align}
\label{eqn:LocalFockBasis}
|n_1,n_2,\dots\rangle=\prod_m\frac{(a_m^\dagger)^{n_m}}{\sqrt{n_m!}}|0\rangle_L, \qquad |\bar n_1,\bar n_2,\dots\rangle=\prod_{m}\frac{(\bar{a}_{m}^\dagger)^{\bar n_{m}}}{\sqrt{\bar n_{m!}}}|0\rangle_L,
\end{align}
We note that the creator and annihilation operators corresponding to different subregions necessarily commute. From this we see that the Fock space built from local modes has a tensor product structure
\begin{equation}
\mf F^L= \mf f\otimes\bar{\mf f},
\end{equation}
where $\mf f$ and $\bar{\mf f}$ are Fock spaces associated with the two regions $[0, r]$ and $[r, R]$. These Fock spaces are defined in the usual fashion by first defining vacuum states $|0\rangle\in\mf f$ and $|\bar 0\rangle\in\bar{\mf f}$ and then the basis states by repeated application of the creators $a_m^\dagger$ and $\bar a_m^\dagger$. For example, the local vacuum for the whole cavity is then the tensor product $|0_L\rangle=|0\rangle\otimes |\bar 0\rangle$ and product states can be written as $|\psi,\phi\rangle=|\psi\rangle\otimes|\phi\rangle$. 

Notice that $|0_L\rangle$ is not a standard vacuum \cite{HaagBook}. Indeed, $|0_L\rangle$ is neither {\em separating} nor {\em cyclic}. It is not separating since $a_l|0_L\rangle=0$ does not imply $a_l=0$. It is not cyclic since it is a product state.
\section{Unitary inequivalence}\label{dealunineq}
So far we have shown that a quantisation based on a different choice of modes, i.e. the local modes,  yields to a different Fock space $\mf F^L$. However, as we shall now see, this Fock space is not unitarily related to the standard $\mf F^G$. 
\subsection{The unitary inequivalence of $\mf F^G$ and $\mf F^L$}\label{subsec:UnitaryInequivalence}
By the sufficient condition for unitary inequivalence stated in Section \ref{sec:BogoliubovTransformations}, all we have to do is to demonstrate that the sum
\begin{align}
\label{eqn:BetaBetadagger}
\sum_m\langle0_G|n_m+\bar n_m|0_G\rangle=\sum_N\langle0_L|N_N|0_L\rangle=\sum_{m,N}\left|(U^*_N|u_m)\right|^2+\left|(U^*_N|\bar u_m)\right|^2,
\end{align}
diverges. To that end it is enough to establish that \eqref{eqn:BetaBetadagger} diverges for each value of $N\in\mathbb N^+$. Explicitly evaluating the sum yields 
\begin{align}
\label{eqn:divergentsum}
\sum_{m}\left|(U^*_N|u_m)\right|^2+\left|(U^*_N|\bar u_m)\right|^2=\sum_{m} \left|\frac{\sin\frac{N\pi r}{R}}{\sqrt{Rr\Omega_N\omega_m}}\frac{\frac{m\pi}{r}}{\Omega_N+\omega_m}\right|^2+\left|\frac{\sin\frac{N\pi r}{R}}{\sqrt{R\bar r\Omega_N\bar\omega_m}}\frac{\frac{m\pi}{\bar r}}{\Omega_N+\bar\omega_m}\right|^2.
\end{align}
We now proceed by making use of the integral test for convergence: the sum diverges iff the corresponding integral diverges. The integral is obtained by simply replacing the index $m$ with a continuous variable $x$, i.e.
\begin{align*}
\int_1^\infty \!\!\!\!dx\!\left(\! \frac{\sin^2\frac{N\pi r}{R}}{Rr\Omega_N\sqrt{\frac{\pi^2x^2}{r^2}+\mu^2}}\frac{\frac{x^2\pi^2}{r^2}}{\left(\Omega_N+\sqrt{\frac{\pi^2x^2}{r^2}+\mu^2}\right)^2}+\frac{\sin^2\frac{N\pi r}{R}}{Rr\Omega_N\sqrt{\frac{\pi^2x^2}{\bar r^2}+\mu^2}}\frac{\frac{x^2\pi^2}{\bar r^2}}{\left(\Omega_N+\sqrt{\frac{\pi^2x^2}{\bar r^2}+\mu^2}\right)^2}\!\right)\!.
\end{align*}
This integrand has the asymptotic behaviour $\sim 1/x$ and therefore \eqref{eqn:divergentsum} diverges, which implies that
\begin{equation}
\sum_N\langle0_L|N_N|0_L\rangle = \langle0_L|N|0_L\rangle =\infty.
\end{equation} 
\subsection{Analysis of the divergences}\label{asymmetry}
In order to proceed, it is important to understand why the sum \eqref{eqn:BetaBetadagger} diverges. As shown in the previous section this behaviour comes from summing over $m$ and not $N$. Specifically, it is easy to show that although summing over $m$ yields an infinite result
\begin{align}
\langle0_L|N_N|0_L\rangle=\sum_{m}\left|(U^*_N|u_m)\right|^2+\left|(U^*_N|\bar u_m)\right|^2=\infty.
\end{align}
The same is not true when summing only over $N$, i.e. we have
\begin{align}
\langle0_G|n_m+\bar n_m|0_G\rangle=\sum_{N}\left|(U^*_N|u_m)\right|^2+\left|(U^*_N|\bar u_m)\right|^2<\infty.
\end{align}
Thus, the global number operators $N_N$ are ill defined in the local Fock space $\mf F^L$, which also implies that $A_N$ and $A_N^\dagger$ are not well-defined operators in $\mf F^L$. Nevertheless, as we shall see in Section \ref{locoponglobfock}, it will turn out that the local number operators $n_m$ and $\bar n_m$ are perfectly well defined in the global Fock space $\mf F^G$. This mathematical asymmetry could be taken as a sign that the global Fock space $\mf F^G$ is in this respect preferred. However, as we shall see below in Section \ref{locevol}, the canonical Hamiltonian \eqref{eqn:GlobalHamiltonian} can be regularised by subtracting the relevant infinite (local) vacuum energy thus rendering the energy expectation values of all basis states in $\mf F^L$ finite and well-defined. Furthermore, we will see in Section \ref{causprop} that states in $\mf F^L$ can be consistently evolved. In this sense, it seems that the unitarily inequivalent global and local quantum field theories are both possible quantizations of the real Klein-Gordon field in the one-dimensional box.
\section{Strictly localised one-particle states in $\mf F^L$ and their causal evolution}\label{locstates}
In this section we shall see that the local quantisation leads to a mathematically meaningful notion of {\em local particles}. We show that these states are strictly localised and that the evolution is causal.
\subsection{Local quanta and their average energy}\label{locevol}
The canonical Hamiltonian $H$ defined by equation \eqref{eqn:GlobalHamiltonian} contains an infinite vacuum energy, which is regularised by subtraction, i.e.  
\begin{align}
H^G=H-\langle0_G|H|0_G\rangle.
\end{align}
This regularised Hamiltonian $H^G$ defines a notion of energy of states in the global Fock space $\mf F^G$. \footnote{Although  the global Hamiltonian $H^G$ is an operator in $\mf F^G$, some states in $\mf F^G$ may  lie outside its domain and thus have an infinite/ill-defined average energy, being for this reason unphysical.}

We now turn to the question of whether we can define a meaningful notion of energy in the local Fock space $\mf F^L$. A good guess is that  the regularised Hamiltonian 
\begin{align}
H^L=H-\langle0_L|H|0_L\rangle,
\end{align}
obtained by subtracting the infinite energy of the local vacuum, is well defined in the local Fock space $\mf F^L$. Let us see how this works out. We first define $\m E$ as the expectation value of $H^G$ on the local vacuum, i.e.
\begin{align}
\m E\equiv \langle 0_L|H^G|0_L\rangle.
\end{align}
Next we compute the energy expectation value of a local $n$-particle state $\langle m_l,\bar 0|H^G|m_l,\bar 0\rangle$. Substituting the Bogoliobov relations
\begin{align}
A_N=\sum_l (U_N|u_l)a_l+(U_N|u_l^*)a_l^\dagger+(U_N|\bar u_l)\bar a_l+(U_N|\bar u_l^*)\bar a_l^\dagger\nn,\\
A_N^\dagger=\sum_l (u_l|U_N)a_l^\dagger+(u_l^*|U_N)a_l+(\bar u_l|U_N)\bar a_l^\dagger+(\bar u_l^*|U_N)\bar a_l,
\end{align}
into the definition of $H^G$ we obtain
\begin{align}
\langle m_l,\bar 0|H^G|m_l,\bar 0\rangle=m_l\sum_N\Omega_N\left(|(u_l|U_N)|^2+|(u_l^*|U_N)|^2\right)+\m E\nn.
\end{align}
For  $m_l=0$ we would have $\langle 0_L|H^G|0_L\rangle=\m E$ and therefore we can write
\begin{align}
\langle m_l,\bar 0|H^L|m_l,\bar 0\rangle=m_l\sum_N\Omega_N\left(|(u_l|U_N)|^2+|(u_l^*|U_N)|^2\right).
\end{align}
From here we see that the local $n$-particle state $|m_l,\bar 0\rangle$ contains $m_l$ units of quanta with the manifestly positive {\em average} energy
\begin{align}
\epsilon_l=\sum_N\Omega_N\left(|(u_l|U_N)|^2+|(u_l^*|U_N)|^2\right).
\end{align}
A simple integral test of convergence reveals that $\epsilon$ is indeed convergent (the corresponding integrand has the asymptotic behaviour $\sim \frac{\sin^2x}{x^2}$). Thus, the regularised Hamiltonian $H^L$ yields finite expectation values for all $n$-particle particle states $|m_l,\bar 0\rangle$. Repeating the above calculations we can also see that the $n$-particle states $|0,\bar n_l\rangle$ have finite energy and so do all basis states $|n_l,\bar n_m\rangle$. Thus, all basis states of $\mf F^L$ and finite superpositions of them will have finite average energy.

\subsection{Strict localisation on the local vacuum}\label{Sec:strictlocpart}
We now proceed to construct {\em strictly localised} one-particle states in $\mf F^L$. As briefly mentioned in the introduction, a  state $|\psi\rangle$ is said to be {\em strictly localised} \cite{Knight:1961} within a region of space $\mf R$ if the expectation value of any local operator $\m O(x)$ outside that region (i.e. $x\notin\mf R$) is identical to that of the vacuum, i.e.
\begin{align}
\langle\psi|\m O(x)|\psi\rangle=\langle0|\m O(x)|0\rangle\,\, \mbox{if} \,x\notin \mf R.\nn
\end{align}
Since we have based our local quantisation on modes $u_m$ and $\bar u_m$ which are localised within the regions $[0, r]$ and $[r, R]$ it is reasonable to expect that the one-particle excitation 
\begin{align}
|1_m,\bar 0\rangle\equiv a_m^\dagger|0_L\rangle=a_m^\dagger|0,\bar 0\rangle\nn,
\end{align}
is strictly localised within $[0, r]$. 

Indeed this is the case. The only operators we can build outside the region $[0, r]$, i.e. in $[r,	 R]$, are expansions in the annihilators and creators $\bar a_m$ and $\bar a_m^\dagger$, and these all commute with $a_m^\dagger$. Hence, we have
\begin{align}
\langle\psi|\m O(\bar a_m,\bar a_m^\dagger)|\psi\rangle&=\langle 0_L|a_m\m O(\bar a_m,\bar a_m^\dagger)a_m^\dagger|0_L\rangle=\langle 0_L|\m O(\bar a_m,\bar a_m^\dagger)a_m a_m^\dagger|0_L\rangle=\langle 0_L|\m O(\bar a_m,\bar a_m^\dagger)|0_L\rangle,\nn
\end{align}
verifying that the state $|1_m,\bar 0\rangle$ is a strictly localised one-particle state. Clearly, the quantisation based on local non-stationary modes provides us with a natural notion of a local particle within the local QFT. Notice however that the notion of strict localisation introduced by Knight in \cite{Knight:1961} made use of the Minkowski vacuum based on stationary solutions of the Klein-Gordon equation. The analogous vacuum state would not be the local vacuum $|0_L\rangle$, but rather the global vacuum $|0_G\rangle$, which is also constructed using stationary modes. As a matter of fact, the possibility of strictly localised states in $\mf F^L$ has to do with the separability of $|0_L\rangle=|0\rangle\otimes|\bar 0\rangle$, a property not shared by  $|0_G\rangle$. Furthermore,  local one-particle states do {\em not} belong to the global Fock space $\mf F^G$, which is, as we have shown above, unitarily inequivalent to $\mf F^L$. We see here that the possibility of local particle states is in our construction intimately related to the existence of unitarily inequivalent representations within QFT.

This construction result should not be considered a mathematical counter-example to the no-go theorems presented in \cite{Knight:1961, Malament, Halvorson}. Indeed, our system does not exhibit translational covariance since we are dealing with a finite box with Dirichlet boundary conditions imposed at the endpoints. It seems nonetheless plausible to us that additional assumptions might be violated in the limit of an infinite unbounded box admitting translation invariance. This possibility should be investigated further.

\subsection{Causal propagation of local states}\label{causprop}
The evolution of states in $\mf F^L$ is defined by the unitary operator $U^L(t)=\exp(-iH^Lt)$ which trivially commutes with $H^L$, implying that the total energy is conserved. We also note that none of the local $n$-particle states are eigenstates of $H^L$, in particular not the local vacuum  $|0_L\rangle$. For this reason it will be interesting to study the evolution of these strictly localised states and verify whether they propagate causally, or not. 

To do this we shall have to introduce a third region $[\tilde r,R]$ with $\tilde r>r$ and the local modes associated with it. We define these modes to be completely localised within $[\tilde r, R]$ at a {\em later} moment in time $t=\tau>0$:
\begin{align}
\tilde u_l(x,t=\tau)&=\frac{\theta(x-\tilde r)}{\sqrt{\tilde r\tilde \omega_l}}\sin\frac{l \pi (x-\tilde r)}{R-\tilde r}=\bar \chi_l(x),\quad \dot{\tilde u}_l(x,t=0)=-i\tilde\omega_l\tilde \chi_l(x)\quad \tilde \omega_l^2=\frac{\pi^2 l^2}{(R-\tilde r)^2}+\mu^2\nn
\end{align}
This defines a new set of creators and annihilators $\tilde a_l$ and $\tilde a_l^\dagger$ related to the global ones as
\begin{align}
\tilde a_l&=\sum_N (\tilde u_l|U_N)A_N+(\tilde u_l|U_N^*)A_N^\dagger\nn,\\
\tilde a_l^\dagger&=\sum_N (U_N|\tilde u_l)A_N^\dagger+(U_N^*|\tilde u_l|)A_N.
\end{align}
The local operators $\tilde{\m O}(\tau)$ associated with the region $[\tilde r, R]$ at time $t=\tau$ will be generated by series expansions in $\tilde a_l$ and $\tilde a_l^\dagger$.

We can now calculate the commutator $[a_m,\tilde a_n^\dagger]$ obtaining
\begin{align}
[\tilde a_n,a_m^\dagger]&=\sum_{M,N}\left[(\tilde u_n|U_N)A_N+(\tilde u_n|U_N^*)A_N^\dagger,(U_M|u_m)A_M^\dagger+(U_M^*|u_m)A_M\right]\nn\\
&=\sum_{M,N}\left[(\tilde u_n|U_N)(U_M|u_m)[A_N,A_M^\dagger]+(\tilde u_n|U_N^*)(U_M^*|u_m)[A_N^\dagger,A_M]\right]\nn\\
&=(\tilde u_n|\left(\sum_{N}|U_N)(U_N-|U_N^*)(U_N^*|\right)|u_m)=(\tilde u_n|u_m)\nn.
\end{align}
An identical calculation yields $[\tilde a_n,a_m]=-(\tilde u_n|u_m^*)$. 

The fact that the local modes propagate causally (see Section \ref{subsec:BogoliubovCoef}) means that $(\tilde  u_n|u_m)$ and $(\tilde u_n|u_m^*)$ are zero whenever $\tau<|r-\tilde r|$, which in turn implies that  $a_m$ and $a_m^\dagger$  commute with $\tilde a_n$ and $\tilde a_n^\dagger$. Thus, any local observable $\tilde{\m O}(\tau)$ will commute with $a_m^\dagger$ and $a_m$ whenever $\tau<|r-\tilde r|$, that is, whenever the spacetime regions associated with the operators $\tilde{\m O}(\tau)$ and the pair $\{a_m,a_n^\dagger\}$ are spacelike. This way, micro-causality is built into the construction.

Besides, we have clearly that
\begin{align}
\langle 1_m,\bar 0|\tilde{\m O}(\tau)|1_m,\bar 0\rangle=\langle 0_L|a_m\tilde{\m O}(\tau)a_m^\dagger|0_L\rangle=\langle 0_L|\tilde{\m O}(\tau)a_ma_m^\dagger|0_L\rangle=\langle 0_L|\tilde{\m O}(\tau)|0_L\rangle,
\end{align}
for $\tau<|r-\tilde r|$, which implies that the local one-particle state $|1_m,0\rangle$ propagates causally as it should. This situation should be contrasted to Knight's strict localisation \cite{Knight:1961} which would state 
\begin{align}
\langle 0_G|a_m\tilde{\m O}(\tau)a_m^\dagger|0_G\rangle =\langle 0_G|\tilde{\m O}(\tau)|0_G\rangle,
\end{align}
which in fact does not hold since, as will become clear below, $a_ma_m^\dagger|0_G\rangle\neq |0_G\rangle$.
\section{Local analysis of the global vacuum}\label{locanal}
In Section \ref{asymmetry}, we pointed to a mathematical asymmetry between the local and global quantum theories. We saw that, while the global number operators are ill defined in $\mf F^L$, the case is different for the local number operators as defined in $\mf F^G$. In this section we shall demonstrate that the local creators and annihilators are indeed well-defined operators in $\mf F^G$, which will allows us to analyse global states using number operators associated with the local quantisation. In particular, we will examine the spectrum of local particles and numerically quantify existent space-like correlations of the global vacuum $|0_G\rangle$.
\subsection{Local operators in $\mf F^G$}\label{locoponglobfock}
Let us now show that the local creator and annihilators $a_l$, $a_l^\dagger$, $\bar a_l$, and $\bar a_l^\dagger$ are well-defined operators in $\mf F^G$. Here we will prove this for $a_l$. The proof is identical for $a_l^\dagger$, $\bar a_l$, and $\bar a_l^\dagger$. 

It suffices to show that $\langle\psi|a_l^\dagger Na_l|\psi\rangle<\infty$ for any basis state $|\psi\rangle=|n_1,n_2,\dots\rangle$ of  $\mf F^G$. We first expand the local annihilator 
\begin{align}
a_l=\sum_N (u_l|U_N)A_N+(u_l|U_N^*)A_N^\dagger.
\end{align}
We have that 
\begin{align}
a_l|n_1,\dots,n_N,\dots\rangle&=\left(\sum_N (u_l|U_N)A_N+(u_l|U_N^*)A_N^\dagger\right)|n_1,\dots,n_N,\dots\rangle\nn\\
&\!\!\!\!=\sum_N (u_l|U_N)\sqrt{n_N}|n_1,\dots,n_N-1,\dots\rangle+(u_l|U_N^*)\sqrt{n_N+1}|n_1,\dots,n_N+1,\dots\rangle\nn.
\end{align}
Multiplying by the number operator $N=\sum_NN_N$, we obtain
\begin{align}
Na_l|n_1,\dots,n_N,\dots\rangle&=\sum_N (u_l|U_N)\sqrt{n_N}(n-1)|n_1,\dots,n_N-1,\dots\rangle\nn\\
&\qquad+(u_l|U_N^*)\sqrt{n_N+1}(n+1)|n_1,\dots,n_N+1,\dots\rangle,
\end{align}
where $n$ is the number of particles of the basis state, i.e. $N|n_1,n_2,\dots\rangle\equiv n|n_1,n_2,\dots\rangle$. Sandwiching with $\langle n_1,\dots,n_N,\dots|a_l^\dagger$ now gives
\begin{align}
\langle n_1,\dots,n_N,\dots|a_l^\dagger N a_l|n_1,\dots,n_N,\dots\rangle&=\sum_N |(u_l|U_N)|^2n_N(n-1)+|(u_l|U_N^*)|^2(n_N+1)(n+1)\nn.
\end{align}
Since
\begin{align}
\sum_N |(u_l|U_N)|^2<\infty,\qquad \sum_N |(u_l|U_N^*)|^2<\infty,
\end{align}
and given that $|n_1,n_2,\dots\rangle$ is a basis state of $\mf F^G$ (therefore satisfying $n=\sum_Nn_N<\infty$), we see that the action of $a_l$ on any basis state is not pathological . An analogous demonstration with minor changes shows that finite expectation values for the global Hamiltonian are also obtained for these vectors.  Thus, since both demonstrations also go  through for $a_l^\dagger$, $\bar a_l$, and $\bar a_l^\dagger$, we have shown that the local creators and annihilators are well-defined linear operators in $\mf F^G$ and gi. Nevertheless, as we have stressed above in Section \ref{asymmetry}, the situation is not symmetric since $A_N$ and $A_N^\dagger$ are not well defined in $\mf F^L$. 
\subsection{Local particle spectrum of the global vacuum}
\begin{figure}[H]
  \centering
    \includegraphics[width=0.7\textwidth]{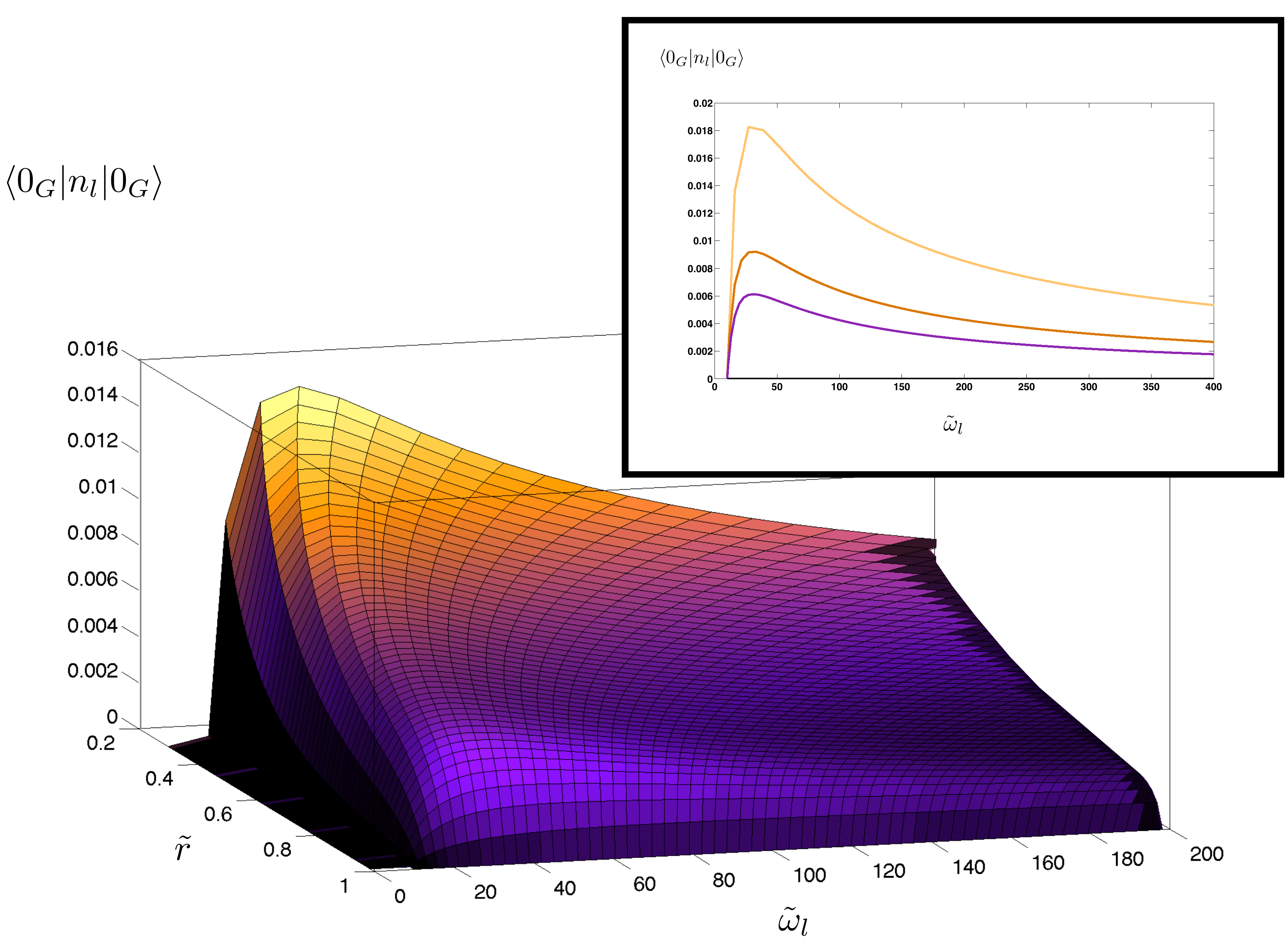}
      \caption{Number of particles for the global vacuum for different sizes of the localisation region $r \in (0.25R,R)$ with fixed $R=1$ and $\lambda_c=\frac{R}{10}$. As expected, when $r=R$ the expectation value of the vacuum is zero for all modes, since local and global modes are the same. }
\label{fig:Rconvergence}
\end{figure}

The global vacuum is defined to have zero global particles, i.e  $\langle0_G|N_N|0_G\rangle=0$. On the other hand, the local quantisation developed above yields a natural notion of local particle number $n_l=a_l^\dagger a_l$, $\bar n_l=\bar a_l^\dagger \bar a_l$, corresponding to the number of local excitations we have in the left and right regions of the box, $[0, r]$ and $[r, R]$, respectively.
\begin{figure}[ht!]
  \centering
    \includegraphics[width=0.7\textwidth]{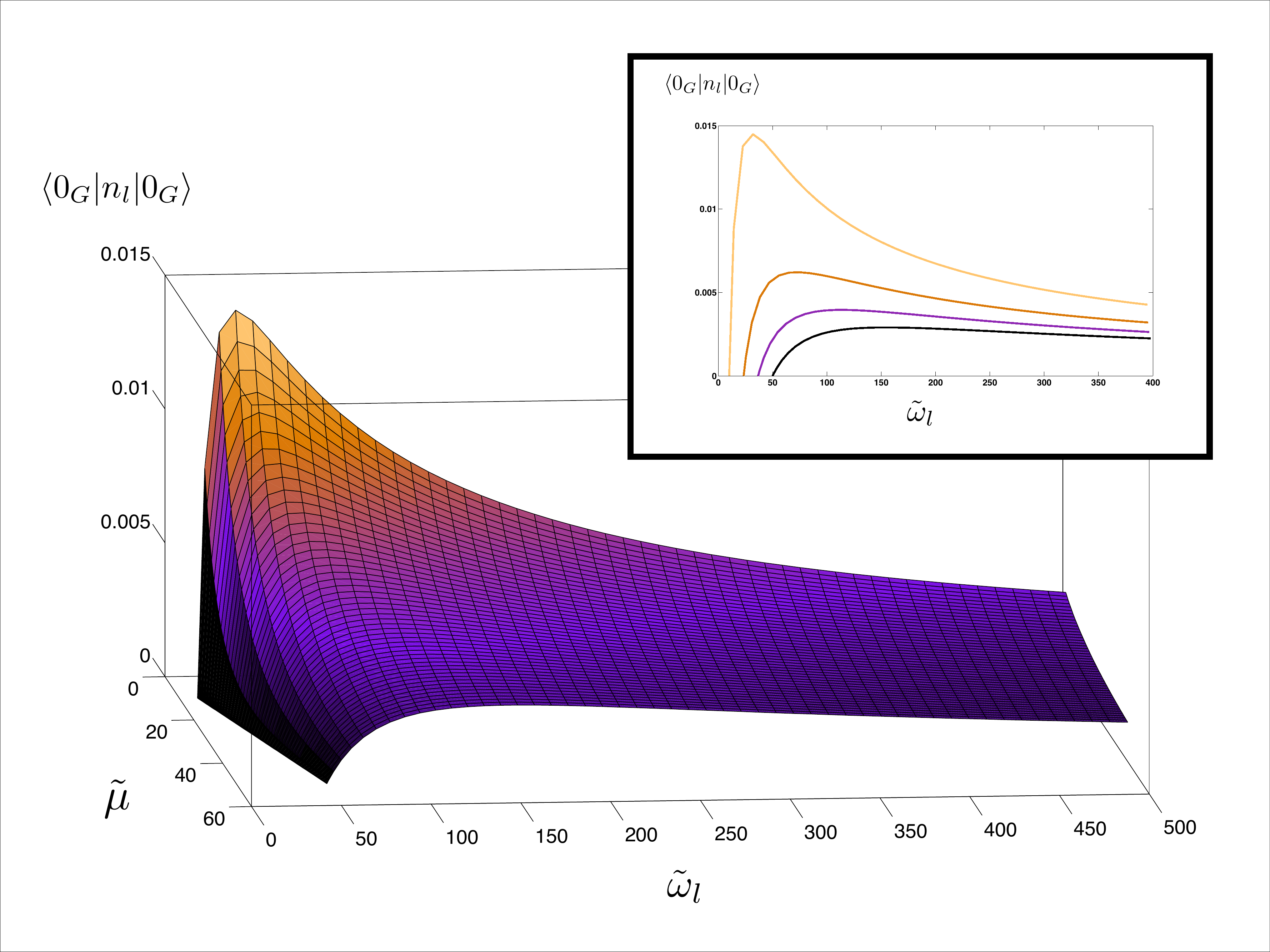}
      \caption{Number of local quanta of energy $\omega_l$ expected value for the global vacuum for different masses $\mu=\frac{\tilde \mu}{R}$ with $\tilde \mu \in (10,50)$. The region of localisation is taken to be $\tilde r^{-1}=R/r=\pi$.  The inset shows discrete values in the same interval for the masses. Higher plots correspond to smaller values. The distribution of local particles in the global vacuum resembles a Planckian spectrum, i.e. a thermal bath of particles.}
\label{fig:Plankian}
\end{figure}
Let us now ask what the distribution of local particles is for the global vacuum. To see this we compute the expectation values 
\begin{align}
\label{eqn:thermalDIst}
 \langle 0_G|n_l|0_G\rangle=\langle 0_G|a^\dagger_l a_l|0_G\rangle =\sum_{N}\frac{l^2\pi^2}{Rr^3\Omega_N\omega_l}\frac{1}{(\Omega_N+\omega_l)^2}\sin^2\frac{\pi N r}{R}.
\end{align}
These depend on three distinct quantities: the size of the cavity $R$, the size of the region of localisation $r<R$, and the mass $\mu$. We could plot the expectation values for different values of these three magnitudes. However, it is more adequate to vary dimensionless quantities, e.g. $r/R$, $r\mu$, and $R\mu$. We might as well fix $R=1$, ending up with two independent dimensionless quantities $\tilde r=r/R$ and $\tilde \mu=R\mu$. Figures \ref{fig:Plankian} and \ref{fig:Rconvergence} show the dependence of the expectation values \eqref{eqn:thermalDIst} on these two variables. 

In Figure \ref{fig:Plankian} we see that when we increase the mass $\mu$ we have that
\begin{align}
\langle0_G|n_l|0_G\rangle=\sum_N|(u_l|U_N^*)|^2\rightarrow0.\nn
\end{align}
In fact, in the large mass limit the coefficients $(u_l|U_N^*)$ have the asymptotic behaviour $\sim\mu^{-2}$ while $(u_l|U_N)$ converge to a non-zero value. Indeed, it is well known that the Compton wavelength $\lambda_C=\mu^{-1}$ determines how well localised a wave-packet, made out of positive frequency modes, can be \cite{NewtonWigner,Compton2,Compton3}. Thus, in the limit $\lambda_C\rightarrow0$, or equivalently $\mu\rightarrow\infty$, the $\beta$-coefficients $(u_l|U_N^*)$ should approach zero.

Another interesting limit is when $r\rightarrow R$, case in which local and global modes converge. Intuitively we would expect  the local description to approach the global one so that the expectation value of local particles goes to zero (since the global vacuum is defined to have zero global particles). This is illustrated in Figure \ref{fig:Rconvergence}. This intuition can be made mathematically precise by studying the convergence of the operators $a_m\rightarrow A_m$ as $r\rightarrow R$. \footnote{We note that for any notion of convergence to make mathematical sense, the operators must act in the same vector space. For example, it is meaningless to claim that $a_k$ converges to $A_N$ as operators defined in the local Fock space $\mf F^L$. Indeed, the operators $A_N$ are not even well defined in $\mf F^L$. Nonetheless, it is meaningful to study the convergence $a_k\rightarrow A_k$ as operators defined in $\mf F^G$.} The relationship between the operators is given by
\begin{align}
a_l=\sum_N(u_l|U_N)A_N+(u_l|U_N^*)A_N^\dagger,
\end{align}
where 
\begin{align}
(u_l|U_N)&=\frac{1}{\sqrt{Rr\Omega_N\omega_l}}\frac{\frac{l\pi}{r}(-1)^l}{\Omega_N-\omega_l}\sin\frac{N\pi r}{R},\nn\\
(u_l|U_N^*)&=-\frac{1}{\sqrt{Rr\Omega_N\omega_l}}\frac{\frac{l\pi}{r}(-1)^l}{\Omega_N+\omega_l}\sin\frac{N\pi r}{R}.
\end{align}
From here it is easy to show that $(u_l|U_N^*)\rightarrow0$ and $(u_l|U_N)\rightarrow\delta_{lN}$ in the limit $r\rightarrow R$. It is now clear that we have convergence of $a_l$ and $A_l$ in the {\em strong} operator topology.

It is important to note that because of unitary inequivalence, the total number of local particles is necessarily infinite, i.e. $\sum_m\langle0_G|n_m+\bar n_m|0_G\rangle=\infty$. In fact, even though Bogoliobov coefficients converge to finite values when $r\rightarrow R$, the sum diverges for any $r$ arbitrarily close to $R$. This is due to the fact that the sum over $m$ and the limit $r \rightarrow R$ do not commute, i.e.
\begin{align}
\label{eqn:limitBetaBetadagger}
\lim_{r\to R} \sum_m\langle0_G|n_m+\bar n_m|0_G\rangle \neq\sum_m \lim_{r\to R}\langle0_G|n_m+\bar n_m|0_G\rangle=0.
\end{align}
Another interesting case is the limit $r\rightarrow0$. Inspecting the coefficients reveal that both $(u_l|U_N)$ and $(u_l|U_N^*)$ have the asymptotic behaviour $\sim r$ and thus vanish  in the limit. However, the sum $\sum_m\langle0_G|n_m|0_G\rangle$ approaches a finite non-zero value when $r\rightarrow0$.  

\subsection{Vacuum entanglement}
As a second application we will look at vacuum entanglement. We shall study the entanglement between the two regions $[0, r]$ and $[r, R]$ by computing the correlations between local particle numbers as given by $\text{cov}(n_m,\bar n_l)$ defined by
\begin{align}
\text{cov}(n_m,\bar n_l)&\equiv\langle \psi|n_m\bar n_n|\psi\rangle-\langle \psi|n_m|\psi\rangle\langle \psi|\bar n_n|\psi\rangle.
\end{align}
We note that if we choose $|\psi\rangle=|0_L\rangle$ then  $\text{cov}(n_n,\bar n_m)$ is identically zero. However, this is not so for the global vacuum $|\psi\rangle=|0_G\rangle$. The correlations of the global vacuum are more conveniently characterised by the dimensionless values
\begin{align}
\text{corr}(n_m,\bar n_n)&=\frac{\langle 0_G|n_m\bar n_n|0_G\rangle-\langle 0_G|n_m|0_G\rangle\langle 0_G|\bar n_n|0_G\rangle}{\sqrt{\langle0_G|n_m^2|0_G\rangle-\langle0_G|n_m|0_G\rangle^2}\sqrt{\langle0|\bar n_n^2|0_G\rangle-\langle0_G|\bar n_n|0_G\rangle^2}}\nn\\
&=\frac{\text{cov}(n_m, \bar n_n)}{\sqrt{\text{cov}(n_m,  n_m) \text{cov}(\bar n_n,  \bar n_n)}},
\end{align}
which are known as the correlation coefficients.
\begin{figure}[H]
  \centering
  \includegraphics[width=\textwidth]{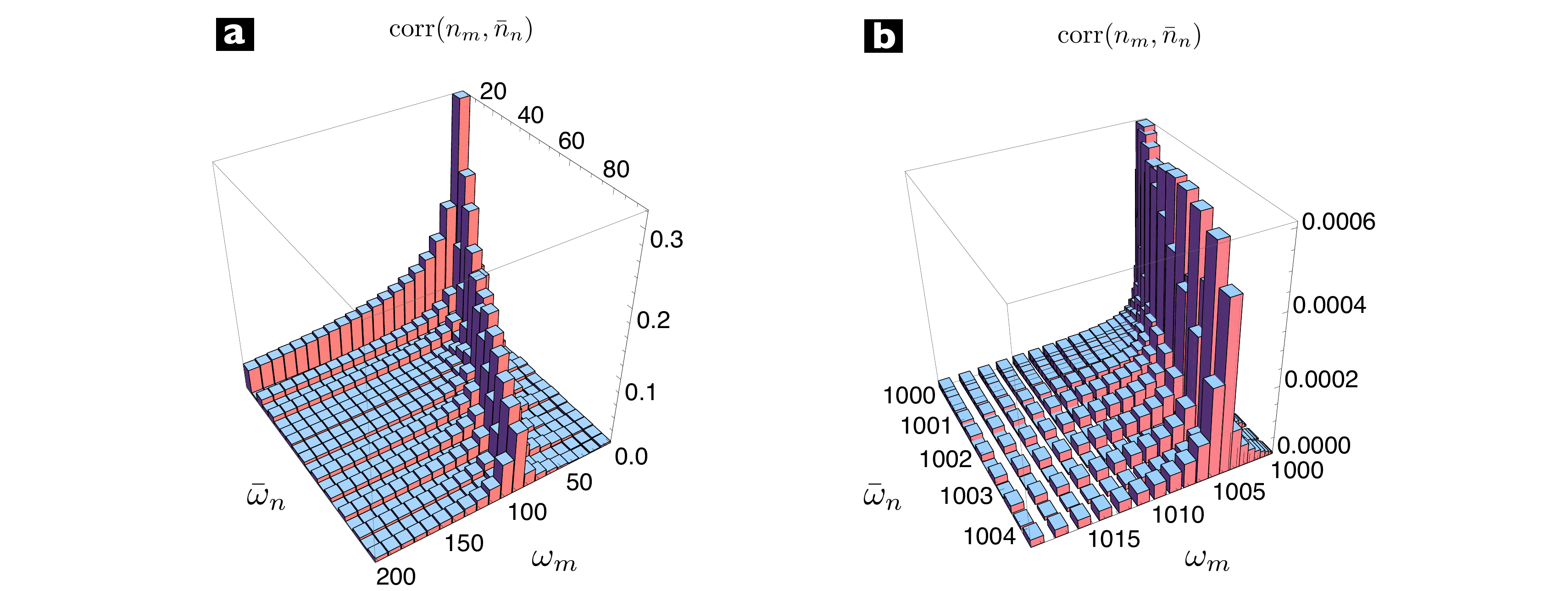}
  \caption{Values for the dimensionless correlation coefficients $\mathcal C(n_m, \bar n_n)$ for the extreme cases of a massless field (a), and highly massive field (b), with $\mu = 1000/R$. For both cases the localisation region $\mf R$ has a size $r= R/\pi$. As we can see, modes with the same frequency are the most correlated ones. Notice that instead of plotting with respect to the mode indexes, we are using the mode frequencies.}
\label{fig:VacuumCorrelations}
\end{figure}
\begin{figure}[H]
  \centering
  \includegraphics[width=\textwidth]{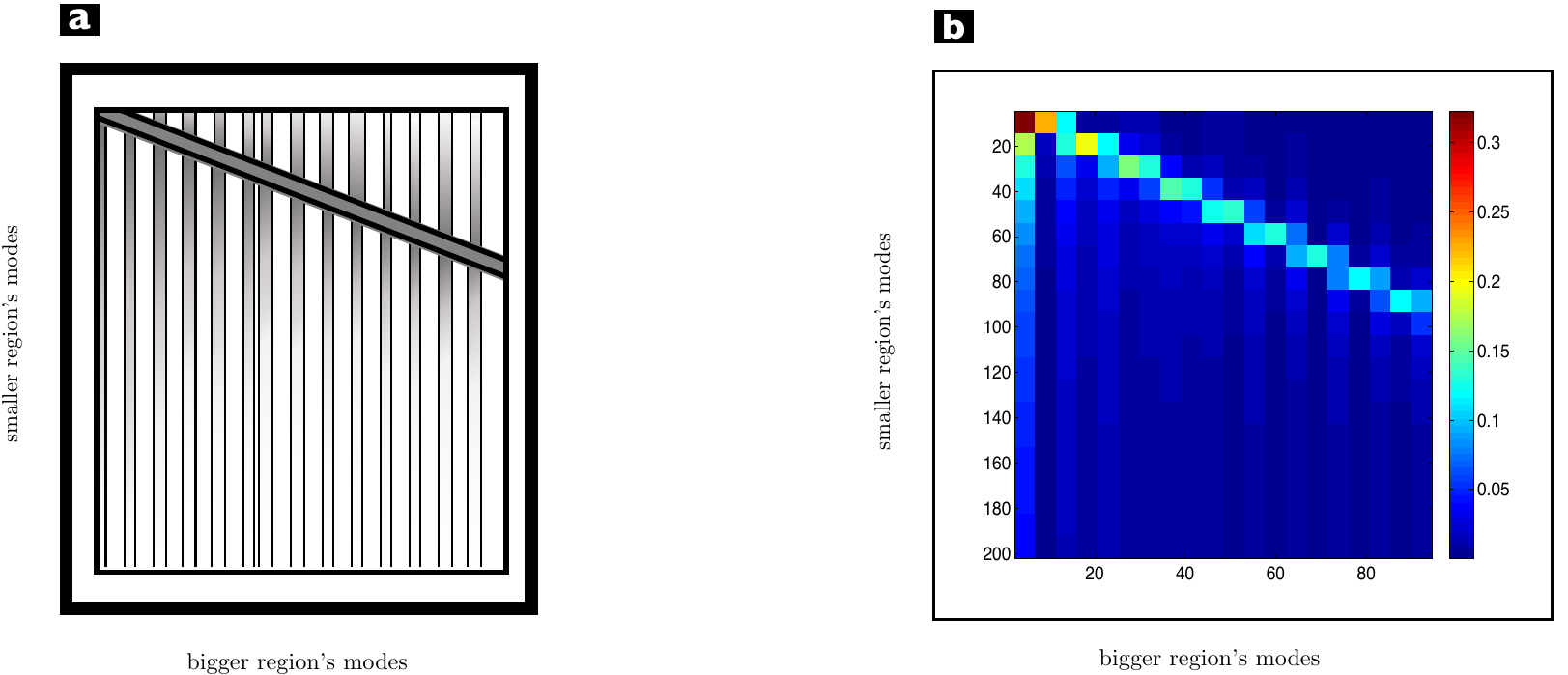}      \caption{(a) A 2D sketch of the features observed in figure \ref{fig:VacuumCorrelations}a. (b) Exact 2D plot of the correlation coefficients as shown in figure \ref{fig:VacuumCorrelations}a.  }
\label{fig:VacuumCorrelationsFeatures}
\end{figure}
From equation \eqref{eqn:afromA} we have 
\begin{multline}
\langle 0_G|n_m\bar n_n|0_G\rangle=\langle 0_G|a_m^\dagger a_m\bar a_n^\dagger\bar a_n|0_G\rangle=\\
=\sum_{M,N}(U_M^*|u_m)(u_m|U_N)(U_N|\bar u_n)(\bar u_n|U_M^*) +(U_M^*|u_m)(u_m|U_N)(U_M|\bar u_n)(\bar u_n|U_N^*)\\
+(U_M^*|u_m)(u_m|U_M^*)(U_N^*|\bar u_n)(\bar u_n|U_N^*)
\end{multline}
On the other hand we have that
\begin{equation}
\langle0_G|n_m|0_G\rangle\langle 0_G|\bar n_n|0_G\rangle=\sum_{M,P}(U_M^*|u_m)(u_m|U_M^*)(U_P^*|\bar u_n)(\bar u_n|U_P^*)
\end{equation}
and thus
\begin{multline}
\langle 0_G|n_m\bar n_n|0_G\rangle-\langle0_G|n_m|0_G\rangle\langle0_G|\bar n_n|0_G\rangle=\\=\sum_{M,P}(U_M^*|u_m)(u_m|U_P)(U_P|\bar u_n)(\bar u_n|U_M^*)+ (U_M^*|u_m)(u_m|U_P)(U_M|\bar u_n)(\bar u_n|U_P^*)
\end{multline}
and using the computed inner products \eqref{eqn:BogoliubovsCoefficients} and equation \eqref{eqn:thermalDIst} we obtain
\begin{align}
\text{corr}(n_m,\bar n_n)&=\frac{\frac{2\pi^4m^2n^2}{R^2r^3\bar r^3\omega_m\bar\omega_n}\sum_{N,P}\left[\frac{(-1)^{N+P}\sin^2\frac{N\pi r}{R}}{\Omega_N\Omega_P(\Omega_N+\omega_m)}\frac{\sin^2\frac{P\pi r}{R}(\Omega_N\Omega_P-\bar\omega_n^2)}{(\Omega_P-\omega_m)(\Omega_N^2-\bar\omega_n^2)(\Omega_P^2-\bar\omega_n^2)}\right]}{\sqrt{\sum_{l,N}\frac{l^2\pi^2}{Rr^3\Omega_N\omega_l}\frac{1}{(\Omega_N+\omega_l)^2}\sin^2\frac{\pi N r}{R}}\sqrt{\sum_{l,N}\frac{l^2\pi^2}{R\bar r^3\Omega_N\bar\omega_l}\frac{1}{(\Omega_N+\bar\omega_l)^2}\sin^2\frac{\pi N \bar r}{R}}}
\end{align}
an expression that can be numerically evaluated, see Figure \ref{fig:VacuumCorrelations}.
Even just a quick look to the figures   \ref{fig:VacuumCorrelations}a and \ref{fig:VacuumCorrelations}b reveals the existence of certain patterns: the extension of correlations along the axis of the small region's local modes (vertical), or the alternance of those extensions (vertical bars) from relevant values to almost zero along the axis of the big region's local modes (horizontal). Although it is out of the scope of this paper to discuss those patterns in detail, we can give a simple explanation of why they would exist, just by thinking in terms of the Fourier decomposition of global modes in terms of small and big local modes (check Figure \ref{fig:cavity}). In order to expand the same global mode, for example $U_{N=1}$, the number of local modes with a relevant contribution will be much higher for the small side than for the big side, the reason for that being, that a \textit{smaller} section of a global mode requires \textit{more} frequencies to be expanded.
So as a matter of fact, those lines also exist along the big region's axis, but they are just much shorter, and so they pass unnoticed. 
Regarding the alternating pattern we can just mention that it has to do mainly with the existence of noticeable differences in the values of the Bogoliubov coefficients for consecutive modes $U_N, U_{N+1}$.

\section{Properties of quasi-local states on $\mf F^G$}\label{sec:quasilocal}
As we have seen in Section \ref{Sec:strictlocpart}, the local quantisation based on non-stationary modes yields a natural notion of local one-particle states in $\mf F^L$ defined by $a_m^\dagger|0_L\rangle$. On the other hand, since the local creators are well-defined in $\mf F^G$, this suggests a natural class of one-particle states $a_m^\dagger|0_G\rangle$ that we will call {\em quasi-local states} defined in $\mf F^G$. In this section we shall examine the properties of these states. In particular, their failure to be strictly localised states is directly related to the Reeh-Schlieder theorem and vacuum entanglement.
\subsection{Positivity of energy}\label{positivenergy}
For historical reasons -- coming from the early attempts of interpreting the solutions of second order Klein-Gordon equation as one-particle wave-functions -- it is commonplace to associate the negative frequency states $U_N^*$ with negative energies, and for this reason to regard them as unphysical states. From that point of view it might seem alarming that we have constructed our local modes using both positive and negative frequency energy-eigenstates, i.e. both $U_M$ and $U_N^*$. Nonetheless, the problem with negative frequencies is a problem in that interpretation and not in relativistic QFT. Indeed, when we adopt the perspective that relativistic QFT arises from the quantisation of a relativistic field, no problems associated with negative frequencies appear. Instead the frequencies are related to energy changes associated with the creation or annihilation of individual quanta.

The classical canonical Hamiltonian
\begin{align}
H=\int dx \frac{1}{2}(\pi^2+(\partial_x\phi)^2+\mu^2\phi^2)\geq0,
\end{align}
being a sum of squares, is manifestly positive definite and is thus bounded from below by zero. As a quantum operator in the corresponding QFT, it is of course ill-defined due to the infinite vacuum energy. Notwithstanding, the regularised Hamiltonian is a sum of the positive operators $N_N$, i.e.
\begin{align}
H^G\equiv H-\langle0_G|H|0_G\rangle=\sum_N\Omega_NN_N.
\end{align}
It is thus clear that any state in $\mf F^G$ has manifestly positive energy and the problem with negative energies is thus avoided by viewing the system, to be quantised, as a classical {\em field} rather than a classical relativistic particle \cite{Strocchi:2004}.

One may be worried that acting with the local creators and annihilators (which were constructed using both positive and negative frequencies) on the global vacuum $|0_G\rangle$, one would obtain unphysical states, perhaps with negative energy. However, as we have demonstrated, the action of the local creators and annihilators on any state $|\psi\rangle\in\mf F^G$ is well defined. Since all states in $\mf F^G$ have manifestly positive energy expectation value it is clear that no problems with negative energy arise.

Nevertheless it is instructive to elaborate on this a bit further. To that end let us investigate whether the state $|\psi_l\rangle=a_l^\dagger|0_G\rangle$ has negative energy. Calculating explicitly the average energy of a state $|\psi_l\rangle=a_l^\dagger|0_G\rangle$, we get 
\begin{align}
\langle\psi_l|H^G|\psi_l\rangle&=\sum_N\Omega_N\langle0_G|a_l N_Na_l^\dagger|0_G\rangle=\sum_{M,N,P}\Omega_N(u_l|U_M)(U_P|u_l)\langle0_G|A_M A_N^\dagger A_NA_P^\dagger|0_G\rangle\nn\\
&=\sum_N\Omega_N(u_l|U_N)(U_N|u_l)=\sum_N\Omega_N|(U_N|u_l)|^2>0,
\end{align}
verifying that the energy is manifestly positive. To demonstrate that the energy is finite we first note that $a_l|0_G\rangle$ is not yet normalised:
\begin{align}
\langle\psi_l|\psi_l\rangle=\langle0_G|a_la_l^\dagger|0_G\rangle=1+\langle0_G|n_l|0_G\rangle\neq1.
\end{align}
The normalised state is therefore given by 
\begin{align}\label{eqn:naivelocpart}
|\psi_l\rangle=\frac{a_l^\dagger|0_G\rangle}{\sqrt{1+\langle0_G|n_l|0_G\rangle}}.
\end{align}
By inspecting the Bogoliubov coefficients \eqref{eqn:BogoliubovsCoefficients} and making use of the integral test of convergence we see that $\langle\psi_l|H^G|\psi_l\rangle<\infty$. Hence, we see that the application of the local creation operator $a_k^\dagger$ on the global vacuum $|0_G\rangle$ keeps the state in the global Fock space $\mf F^G$, i.e. $|\psi\rangle\in\mathfrak F^G$. 

We can also consider the state
\begin{align}
|\phi_l\rangle=\frac{a_l|0_G\rangle}{\sqrt{\langle0_G|n_l|0_G\rangle}},
\end{align}
which is not zero since $a_l$ contains both $A_N$ and $A_N^\dagger$, nor does it have less energy than the global vacuum state. A calculation similar to the one above shows that the energy is manifestly positive $\langle\phi|H|\phi\rangle>0$. Again by the integral test of convergence we could check that the state has, in fact, a finite energy expectation value.

\subsection{Quantum steering and the Reeh-Schlieder theorem}
We are now in a position to address the question of whether the normalised state
\begin{align}
|\psi_m\rangle=\frac{a_m^\dagger|0_G\rangle}{\sqrt{1+\langle0_G|n_m|0_G\rangle}},
\end{align}
can be viewed as a strictly localised one-particle state. The associated wave-packet defined by $\psi_m(x,t)\equiv\langle0_G|\phi(x,t)|\psi_m\rangle$ is in fact the positive frequency part of $u_m$, defined in \eqref{modexp}. One might naively suspect that these states should be localised states since they are created by a {\em local} operation on the vacuum state, i.e. $|0_G\rangle\rightarrow a_m^\dagger|0_G\rangle$. The components of this state in the global basis \eqref{eqn:globfockbasis} are given by
\begin{align}
\frac{a_m^\dagger|0_G\rangle}{\sqrt{1+\langle0_G|n_m|0_G\rangle}}=\frac{\sum_N (U_N|u_m)A_N^\dagger+(U_N^*|u_m)A_N|0_G\rangle}{\sqrt{1+\langle0_G|n_m|0_G\rangle}}=\frac{\sum_N (U_N|u_m)|1_N\rangle}{\sqrt{1+\langle0_G|n_m|0_G\rangle}},
\end{align}
which we recognise as a superposition of global one-particle excitations. From an analysis by Knight \cite{Knight:1961} showing that no finite superposition of $N$-particle states can be strictly localised, we already know that $|\psi_m\rangle$ is not strictly localised. We could stop here, but it is interesting to gain more understanding why this happens.

To investigate this fact, let us see whether the expectation value $\langle\psi_m|\bar n_l|\psi_m\rangle$ is different from $\langle0_G|\bar n_l|0_G\rangle$. Computing this difference yields
\begin{align}
\langle\psi_m|\bar n_l|\psi_m\rangle-\langle0_G|\bar n_l|0_G\rangle&=\frac{\langle0_G|a_m\bar n_l a_m^\dagger|0_G\rangle}{1+\langle0_G|n_m|0_G\rangle}-\langle0_G|\bar n_l|0_G\rangle,\nn\\
&=\frac{\langle0_G|n_m\bar n_l|0_G\rangle-\langle0_G|\bar n_l|0_G\rangle\langle0_G|n_m|0_G\rangle}{1+\langle0_G|n_m|0_G\rangle}\propto \text{corr}(n_m,\bar n_l),
\end{align}
which not only shows that the one-particle state $|\psi_m\rangle$ is not strictly localised, but also tells us that the reason for it is vacuum entanglement. Indeed, making the replacement $|0_G\rangle\rightarrow|0_L\rangle$ and $|\psi_m\rangle\rightarrow |1_m,\bar 0\rangle$ we have $\text{corr}(n_m,\bar n_l)=0$ and the above difference disappears.

It may seem puzzling that we can change the expectation values in the region $[r, R]$ by performing a {\em local} operation in $[0, r]$. Does this not imply the possibility of superluminal signaling? The answer is no, the reason being that the operation $|0_G\rangle\rightarrow |\psi_m\rangle$ is not a unitary operation on the vacuum state since $a_m a_m^\dagger\neq 1$. This local operation does not correspond to something which can be achieved physically by local manipulations solely in $[0, r]$. However, with suitable post-selection, the operation $|0_G\rangle\rightarrow |\psi_m\rangle$ could perhaps be implemented, but only by informing the observer in the region $[r, R]$ which states to post-select. This of course would require classical communication, limited by the speed of light \cite{Summers:1985}.

We can view this in the context of the Reeh-Schlieder theorem \cite{Reeh:1961}. This theorem states that by a local {\em non-unitary} operation in a finite region in space we can obtain, to arbitrary precision, any state at a spatially separated region. The theorem does not go through if we restrict ourselves to local unitary operations. 

The situation is different when we replace the global vacuum $|0_G\rangle$ with the local vacuum $|0_L\rangle$. As seen in Section \ref{Sec:strictlocpart} the key difference is that the local vacuum $|0_L\rangle$ neither cyclic nor separating, or more simply, it is a product state $|0_L\rangle=|0\rangle\otimes|\bar 0\rangle$ which is therefore not entangled. Thus, no steering whatsoever could take place in this case.

\subsection{Further properties}
In the section \ref{positivenergy} we analysed the positivity of energy of the pseudo-local states 
\begin{align}
|\psi_l\rangle=\frac{1}{\sqrt{1+\langle0_G|n_l|0_G\rangle}}a_l^\dagger|0_G\rangle\qquad |\phi_l\rangle=\frac{1}{\sqrt{\langle0_G|n_l|0_G\rangle}} a_l|0_G\rangle,
\end{align}
which are in fact superpositions of global one-particle states $|1_N\rangle=A_N^\dagger|0_G\rangle$, i.e.
\begin{align}
|\psi_l\rangle=\frac{1}{\sqrt{1+\langle0_G|n_l|0_G\rangle}}\sum_N(u_l|U_N)|1_N\rangle\qquad |\phi_l\rangle=\frac{1}{\sqrt{\langle0_G|n_l|0_G\rangle}} \sum_N(u_l|U_N^*)|1_N\rangle
\end{align}

\begin{figure}[H]
  \centering
  \includegraphics[width=0.7\textwidth]{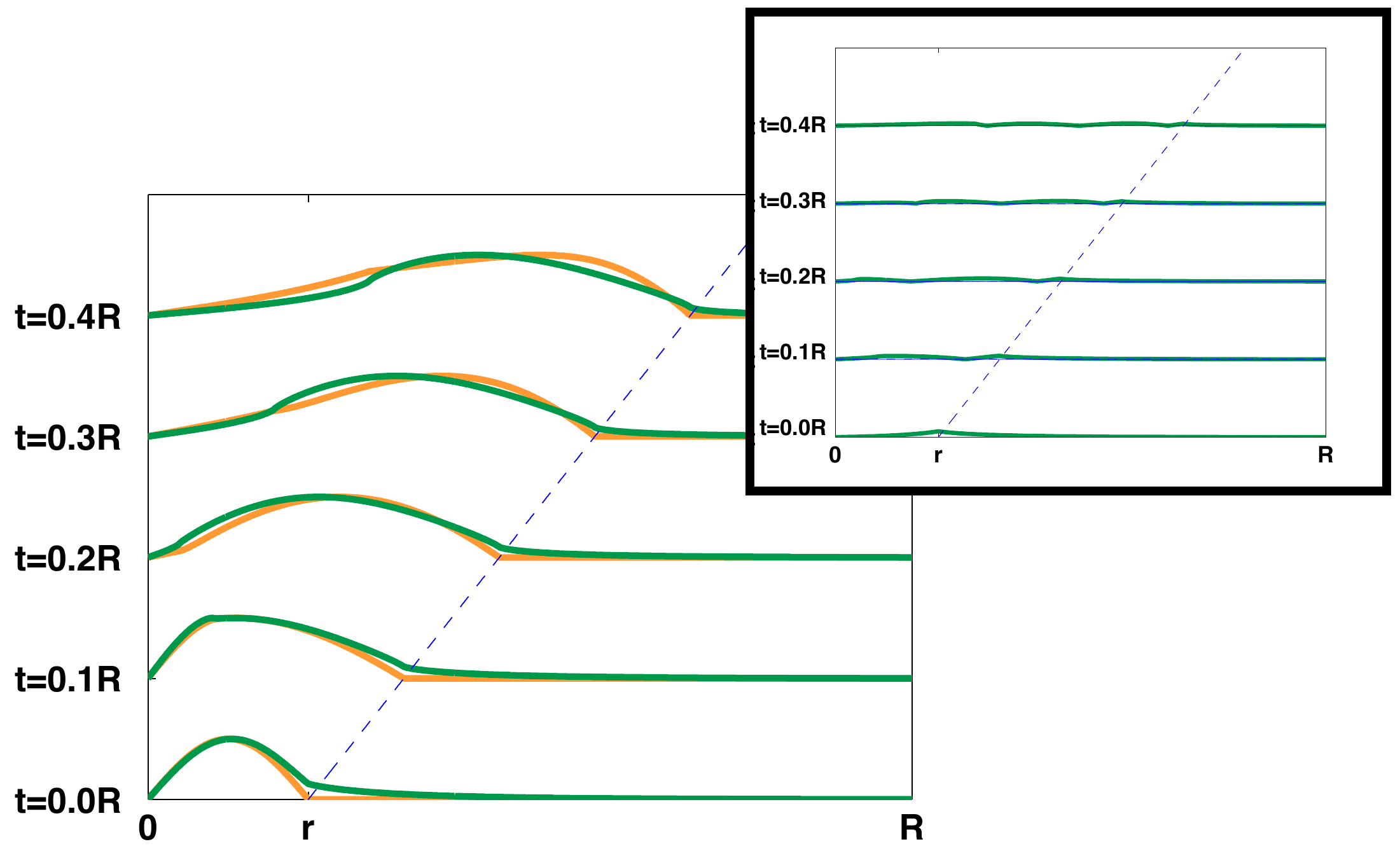}      \caption{Quasi-local modes as compared with local modes. The picture shows the particular case of $r=0.21R$, $\mu= 1/r$ with mode number $m=1$.  It portraits local modes (zero valued out of the light cone) and quasi-local modes, showing exponential decaying fall-offs around the light cone. The inset shows the difference of both modes at the same scale.}
\label{fig:quasiloc}
\end{figure}

Let's define $|\psi^{(r)}_l\rangle = a^{(r)\dagger}_l|0_G\rangle$, where the $(r)$ superindex refers to the operator corresponding to a localisation region of size $r$. We would expect that state to resemble a one-particle local state, in the sense that the corresponding mode would just  be the positive frequency part of the one-particle local mode. That is indeed the case. Figure \ref{fig:quasiloc} illustrates this case for a particular case of those shown in figure \ref{fig:UEvolution}. We would therefore call these modes, which lie in the global Fock space $\mf F^G$, quasi-local modes. 
For all practical purposes this kind of states could be used as localised and causal to a very good approximation. 

Besides that, it is interesting to study how much $ | \psi^{(r)}_l\rangle$  states resemble to the one-particle global states, and therefore we will calculate the expectation value : 

\begin{equation} 
\langle \psi^{(r)}_l | A^{\dagger}_N A_N | \psi^{(r)}_l\rangle
\end{equation}

which happens to be identically equal to

\begin{equation} 
|\langle 1_N | \psi^{(r)}_l\rangle|^2 = |\langle 0_G| A_N | \psi^{(r)}_l\rangle|^2 =  \frac{|\langle 0_G| A_N a^{(r)\dagger}_l  | 0_G\rangle|^2}{1+\langle0_G|n_l|0_G\rangle} = \frac{|(U_N|u_l)|^2}{1+\langle0_G|n_l|0_G\rangle} 
\end{equation}

\begin{figure}[H]
  \centering
  \includegraphics[width=\textwidth]{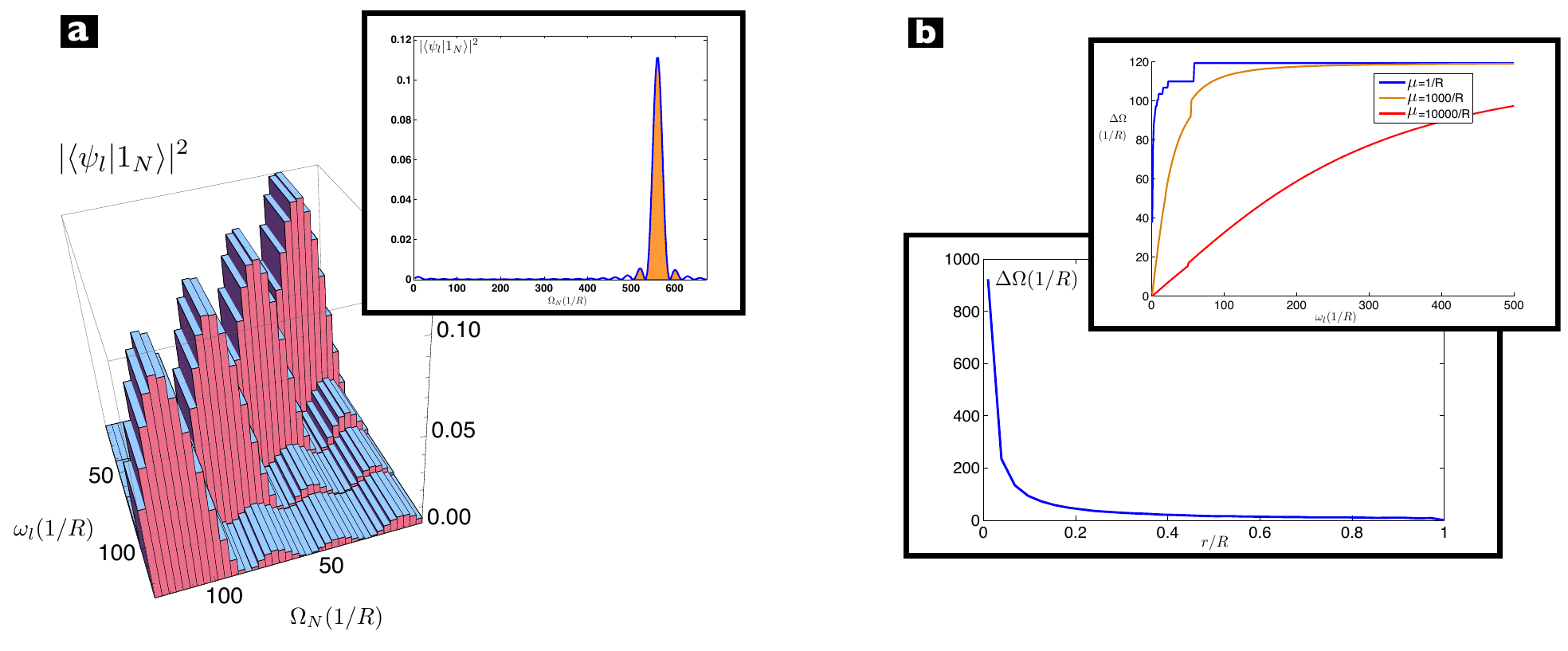}      \caption{Quasi-local state analysis. a) Decomposition of $|\psi_l\rangle$ states in terms of $|1_N\rangle$ global states for a massless case with r=R/9. In the inset, the particular case for $l=20, \omega_l = \pi l /r \simeq 571/R$. b) The estimated bandwidth $\Delta \Omega$ for quasi-local states is independent of the mode $l$ for big $l$, but shows a strong dependence with $r/R$. The inset shows the dependence of  $\Delta \Omega$ with $l$ for different masses $\mu$ for the case  $r=R/9$. }
\label{fig:overlapglocal}
\end{figure}

Figure \ref{fig:overlapglocal}a shows the expansion of $| \psi^{(r)}_l\rangle$ in terms of global particle states  $| 1_N\rangle$ for the massless case. We can see that the decomposition is a rather peaked one, and in particular, we can estimate a bandwidth $\Delta\Omega$ for the expansion in global modes. We can define it as the smallest $\Delta \Omega$ for which:

\begin{equation} 
\sum_{\Omega_N \in (\omega_l- \Delta\Omega/2,\omega_l+\Delta\Omega/2)} |\langle 1_{N} | \psi^{(r)}_l\rangle|^2 > 0.95
\end{equation}

In the general case, $\Delta \Omega$  depends on the frequency of the mode $\omega_l$, but tends to an asymptotic value in the limit of big $l$'s, as we can see in the inset of Figure \ref{fig:overlapglocal}b, where the dependence with the Klein Gordon mass $\mu$ is also plotted. The asymptotic value is independent on the mass, and only dependent on the $r/R$ value. The relationship between these two can be seen in Figure \ref{fig:overlapglocal}b.  In the limit of small values of $r/R$, which would correspond to strongly ``localised particles'', the bandwidth tends to infinity, i.e. we need an infinite amount of global modes to describe the quasi-local particle. For high values of $r/R$ the bandwidth approaches a minimum and we can approximately identify the quasi-local particle states with global states. 

\section{Conclusions and outlook}
In the extant literature there are several theorems and results that indicate the impossibility of having local particle states, e.g.  \cite{Malament,Halvorson,Hegerfeldt:1998,Knight:1961}. We believe that the main obstruction comes from postulating that the one-particle Hilbert space is spanned by positive frequency modes. In particular, no wave-packet built from these modes can be localized within a finite spatial region, even for an arbitrarily small time interval. However, as pointed out in Wald's exposition of the quantization procedure \cite{WaldCurvedBook}, there is nothing preventing us from making use of a different set of modes. The basic idea of this paper was that, basing the quantization procedure on localized modes, we might account for localized one-particle states. Indeed, this turns out to be the case.

These local modes are defined by their initial data. Both the value and time-derivative of the modes are taken to be completely localized within either the right or left partition of the box. This data defines a well-posed Cauchy problem. By Hegerfeldt's theorem, these solutions of the Cauchy problem must contain both positive and negative frequency modes. This marks, at the classical level, a point of departure from the standard quantization procedure. 

The creation and annihilation operators associated with these local modes can then be used to build a Fock space $\mf F^L$, whose basis states describe local elementary excitations of the quantum field. A set of these basis states, e.g. $|n_k,\bar 0\rangle$, does in fact represent strictly localized states with respect to the local vacuum $|0_L\rangle\in\mf F^L$. This vacuum state, however, does not share the typical properties of a quantum field vacuum. In particular, it is neither cyclic not separating, as it is free from correlations between left and right partitions.

Intriguingly, the local and standard (global) quantum field theories turn out to be {\em unitarily inequivalent}. Specifically, by computing the Bogoliobov coefficients relating the global and local quantum theories we have found that 
\begin{align}\label{unequiv}
Tr\ \beta^\dagger \beta \equiv\sum_{k,N}|(u_k|U_N)^*|^2+|(\bar u_k|U_N^*)|^2, 
\end{align}
diverges, which is a sufficient condition for establishing unitary inequivalence. Nevertheless, it is important to note that both standard and local quantizations produce self-consistent quantum theories of the field. As a matter of fact, as we have demonstrated, we can evolve states and we also have a well-defined notion of energy after the local vacuum energy has been subtracted from the canonical Hamiltonian. 

The existence of unitarily inequivalent representations would seem to confront us with a problem of which Fock space representation to choose \cite{Fraser:2011}. The problem of unitary inequivalence disappears, however, when some form of regularisation is introduced \cite{Wallace:2006}. Imposing of a wave-number $k=\pi m/r$ cutoff, for example, could solve the issue. Such a cut-off would come naturally, for example, from a quantum theory of gravity requiring a discretisation of space(time). A restatement of the theory, which considers the use of measurement apparatuses for a finite time, would also imply the introduction of a frequency cut-off, circumventing the divergences present in \eqref{eqn:divergentsum}. 

Within our approach we nevertheless find that there is a mathematical asymmetry between the two Fock space representations. In fact,  the divergence of the sum \eqref{unequiv} originates from the summation over the local-mode numbers $m$. On the other hand, the sum over global-mode numbers $N$ is finite for each specific value of $m$. A consequence of this fact is that the local creators and annihilators are well-defined operators in the global Fock space, and so are the local number operators. However, the global creators and annihilators turn out to be ill-defined on $\mf F^L$. This asymmetry could perhaps be taken as an indication that the global Fock space representation is preferred. 

In any case, the fact that both local creators and annihilators are well-defined in $\mf F^G$ provides us with a useful set of mathematical tools to analyse the properties of the states in $\mf F^G$. In particular, by computing the expectation values of the local number operators, we have shown that the global vacuum $|0_G\rangle$ is characterised by a bath of local particles. We also showed, by calculating the correlation coefficients of local number operators, that the local particles associated with the left and right regions are highly entangled in the global vacuum, a feature not shared by the local vacuum $|0_L\rangle$.

Again, the well-defined character of local creators and annihilators in $\mf F^G$ also allows us to introduce a new set of quasi-local states defined by applying the local creation operator on the global vacuum, i.e. $|\psi_m\rangle\sim a_k^\dagger|0_G\rangle$. These are natural candidates for {\em essentially localized} states \cite{Haag:1965}. We have also shown how these states fail to be strictly localized, a fact related to vacuum entanglement and the Reeh-Schieder theorem.

Unitary inequivalence seems to be the key problem in the construction of particle localised states, and that could connect with the abstract no-go results by Malament \cite{Malament} and Clifton {\em et. al} \cite{Halvorson}. However, a proper analysis of this matter would require an adaptation of our setup to incorporate translation covariance,  which is an essential assumption in the theorems mentioned.

Clearly there are several topics that deserve further exploration. Here we mention a few of them. For example, it would be nice to express the global vacuum state using the eigenstates $|n_k,\bar n_l\rangle$ of the local number operators.\footnote{Although the local number operators $n_m$ and $\bar n_l$ are well-defined Hermitian operators in $\mf F^G$ we note that their eigenstates $|n_m,\bar n_l\rangle$ belong to $\mf F^L$ and {\em not} to $\mf F^G$. The situation is similar for a non-relativistic quantum particle in a box where the eigenstates of the self-adjoint momentum operator $p=-i\partial_x$ do not belong to the Hilbert space because they do not satisfy the Dirichlet conditions at the boundary.} Such an expression would allow us to construct the reduced density matrix for the regions $[0, r]$ and  $[r, R]$ by partial tracing. From there it would be interesting to see whether the reduced state takes the form of a KMS state. Hopefully we could make contact with existing literature, which examines the entanglement and thermality connected to localised regions of space \cite{Haggard:2013}. In that respect it is perhaps interesting to note that our construction, in contrast to the Minkowski and Rindler quantizations, was not based on standard stationary states. Indeed, while the Minkowski and Rindler quantizations both rely on stationary modes with respect to  time translation and boost operators respectively, our construction makes use of manifestly non-stationary states. Whether this provides some advantage remains to be seen. In any case, it would be of interest to analyse in detail the differences and similarities between the Rindler quantisation and the one presented in this paper.

\vspace{0.5cm}
\noindent{\bf Acknowledgements:} We have benefited from discussions with Luis Garay, Guillermo A. Mena Marug\'an, Juan Manuel P\'erez-Pardo, Hans Halvorson and Jakob Yngvason. H. Westman is grateful for initial discussions with Fay Dowker on the possibility of having local qubits in QFT. This work is supported by Spanish MICINN Projects FIS2011-29287 and CAM research consortium QUITEMAD S2009-ESP-1594. M. del Rey was supported by a CSIC JAE-PREDOC grant. H. Westman was supported by the CSIC JAE-DOC 2011 program. 

\bibliographystyle{unsrt} 



\end{document}